\documentclass[lettersize]{IEEEtran}
\usepackage{amsmath,amsfonts}
\usepackage{algorithmic}
\usepackage{algorithm}
\usepackage{array}
\usepackage[caption=false,font=normalsize,labelfont=sf,textfont=sf]{subfig}
\usepackage{textcomp}
\usepackage{stfloats}
\usepackage{url}
\usepackage{verbatim}
\usepackage{graphicx}
\usepackage{cite}
\begin{document}
	
	\title{3D Electronic-Photonic Heterogenous Interconnect Platforms\\Enabling Energy-Efficient Scalable Architectures\\For Future HPC Systems}
	
	\author{Anirban Samanta, Shun-Hung Lee, Chun-Yi Cheng, Samuel Palermo,~\IEEEmembership{Senior~Member,~IEEE}, and S.~J.~Ben~Yoo,~\IEEEmembership{Fellow,~IEEE,~Optica}% <-this % stops a space
        \thanks{Anirban Samanta, Shun-Hung Lee, and S. J. Ben Yoo are with the Department of Electrical and Computer Engineering, University of California, Davis, CA, USA}% <-this % stops a space
        \thanks{Chun-Yi Cheng, and Samuel Palermo are with the Analog and Mixed-Signal Center, Texas A$\&$M University, College Station, TX, USA, (e-mail: pchang0628@tamu.edu)}% <-this % stops a space
		\thanks{Manuscript received April 19, 2021; revised August 16, 2021.}}

	\IEEEpubid{0000--0000/00\$00.00~\copyright~2021 IEEE}
	
	\maketitle
	\begin{abstract}
		3D interconnects have emerged as a solution to address the scaling issues of interconnect bandwidth and the memory wall problem in high-performance computing (HPC), such as High-Bandwidth Memory (HBM). However, the copper-based electrical interconnect retains fundamental limitations. Dense I/O for high-speed signals lead to degraded signal quality for end-to-end links, necessitating additional circuits to mitigate signal impairments and resulting in poor energy efficiency. We propose a 3D chiplet stacking electronic-photonic interconnect (EPIC) platform, which offers a solution by moving the high-speed data communication interface to the optical domain across the 3D stack by using Through Silicon Optical Vias (TSOV), while retaining the functionality of electrical TSVs and 2.5D interconnects for power delivery and short-reach low-latency communications. We then benchmark the proposed model against state-of-the-art 3D electrical interconnects to demonstrate our 3D EPIC platform beating the 3D electrical interconnects to $>$10\ TB/s/$mm^2$ bandwidth density. We present a pathway to extend our demonstrated, industry-ready design to achieving $\leq$100 fJ/bit high-speed communication.
	\end{abstract}
	\section{Introduction}	
	\IEEEPARstart{H}{igh} -performance computing (HPC) demands have increased massively during the past decade, due to the emergence of new applications of machine learning (ML) and generative Artificial Intelligence (AI) models. Since 2010, the compute needs of training generative AI (genAI) models have increased several orders of magnitude from needing petaFLOPs ($10^{15}$) to requiring yottaFLOPs ($10^{24}$) today for training the latest models to projected quettaFLOPs ($10^{30}$) by 2030 \cite{epoch_models}. With data movement being the primary source of energy cost, traditional copper-based electrical interconnects are a major bottleneck for energy-efficient computing systems due to fundamental limitations of signal integrity challenges in dense high-speed communication. Beyond extreme short-reach (XSR) links, high-speed communication requires expensive digital signal processing circuits to maintain signal integrity, which is compounded by the massively parallel dense I/O requirements of modern shoreline-limited chips. 3D-IC architectures, such as 3D-stacked HBM memory, have emerged in recent years; however, by utilizing copper interconnects for communication, they retain the same fundamental limitations. Photonic interconnect adoption is on the rise, with several startups, including Lightmatter, Celestial.AI, and Ayar Labs, pursuing photonic interconnect platforms to overcome the limitations of copper. The reticle-sized platforms require wafer-scale fabrication, making them costly and difficult to scale. There is an opportunity if smaller, high-yield, node-optimized chiplets can be co-integrated and 3D-stacked together on a single platform, which also resolves the high-speed communication I/O limitations of copper-based electrical interconnects. A 3D electronic-photonic interconnect platform on an active optical interposer featuring vertical optical channels with Through Silicon Optical Vias (TSOV) can be a solution by bringing a global optical interconnect to every high-speed communication node directly in a 3D-Chiplet stack for a new paradigm of 3D-Chiplet stacked architectures, while retaining traditional TSVs, it maintains compatibility with monolithic control circuits, other heterogeneous ICs, and supports easier design of power delivery network.
    \IEEEpubidadjcol
	
    \section{Trends, and Performance Metrics}
	Scalability in AI HPC is critically power-limited; however, notable constraints also exist in fabrication, memory capacity, and memory access latency, as shown by \cite{epoch_scaling_2030}. Overcoming these limitations and meeting performance targets will require scaling every system metric, including transistor count, density, input/output (I/O) density, and, most importantly, energy efficiency \cite{epoch_scaling_2030}. 
	\begin{figure*}
		\centering
		\includegraphics[width=6.0in]{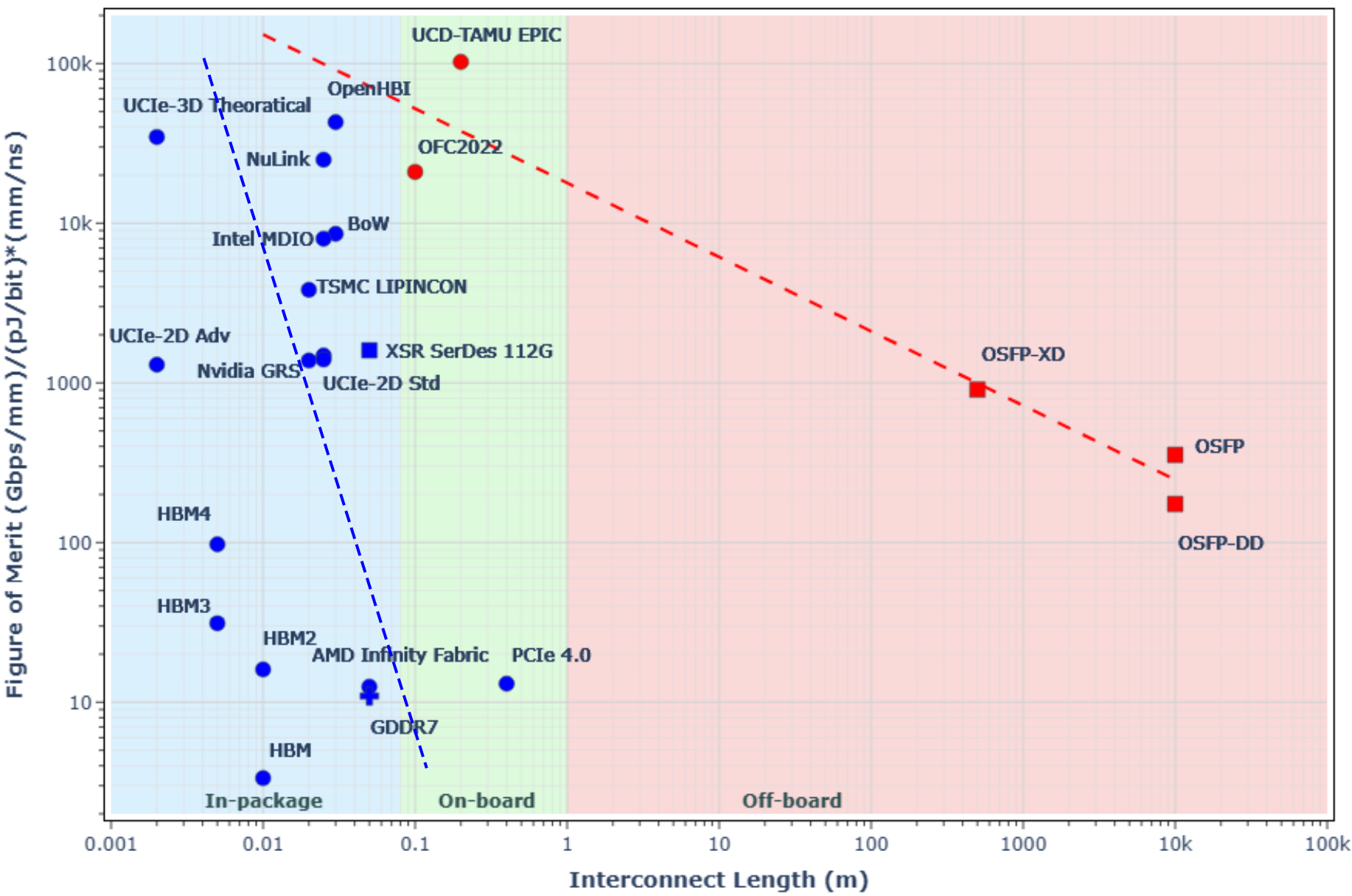}
		\caption{Comparison of interconnect technologies performance Figure of Merit (FoM) (Gbps/mm)/(pJ/bit)*(mm/ns) for the link length or reach of a typical implementation.}
		\label{fom_metric}		
	\end{figure*}	
    Physical limitations such as die/package size, available shoreline for the interconnect I/O, and pocket power budget have become bottlenecks for scaling modern microprocessor architectures, leading to issues such as the "memory wall" problem \cite{gholami2024}. This led to the development of 3DIC architectures, such as HBM \cite{kim2021,zhou-micron-2023,kim2024}, chiplets \cite{naffziger-AMD-2021,loh-AMD-2023}, hybrid bonding \cite{elsherbini-intel-2021}, and dense interconnects, including Intel EMIB and UCIe \cite{duang2021,ucie2024}, to overcome these bottlenecks. Continued scaling poses substantial challenges in the conventional electrical domain, as the copper interconnect itself is a limiting factor. Focus on performance scaling has moved to the interconnect from the on-die transistors, given that the energy cost of data movement over the interconnects is orders of magnitude higher than the energy cost of computing \cite{dally2011}, and has been shown to account for 62.7\% of the total power budget in some studies \cite{ghose2018}. This is due to the long copper interconnects used for high-speed data transmission suffering from several limitations: signal attenuation resulting from conductive and dielectric losses, crosstalk-induced noise, dispersion-induced jitter, and electromagnetic interference, which severely degrade signal integrity and must be taken into account when considering interconnect performance. The overall interconnect performance must be characterized in terms of performance metrics and a Figure of Merit (FoM) that accounts for energy efficiency, bandwidth density, silicon efficiency, and link latency. Silicon efficiency refers to the silicon area required for implementing the physical layer (PHY) transceiver circuit. The bandwidth density and silicon efficiency can be combined into a bandwidth efficiency factor, measuring the ratio of the areal bandwidth density achieved over the transceiver die's shoreline bandwidth density, as these factors are often reported in literature. This measures how efficiently the silicon space is utilized to meet bandwidth targets. An earlier FoM proposed by \cite{keeler2018}, leaves out the link latency from the metric itself. As high-speed electrical links need to trade off bandwidth density for link reach, given signal integrity demands, the inclusion of link latency is crucial. We plot the FoM over a typical implementation for several commercial and in-research interconnect technologies in Fig.\ref{fom_metric}. For electrical interconnects, only Complementary Metal-Oxide-Semiconductor (CMOS) based technologies are considered, as it's the predominant option in low-power digital electronics. Some metrics are not publicly available, so a best estimate is made based on the published literature.
    \begin{equation} \label{eq:fom}
    	\text{FoM} = 
    	\frac{\text{Bandwidth Efficiency}}{\text{Energy Efficiency}} \times \frac{\text{Link Length}}{\text{Link Latency}}
    \end{equation}
    \begin{figure*}
		\centering
		\includegraphics[width=7.0in]{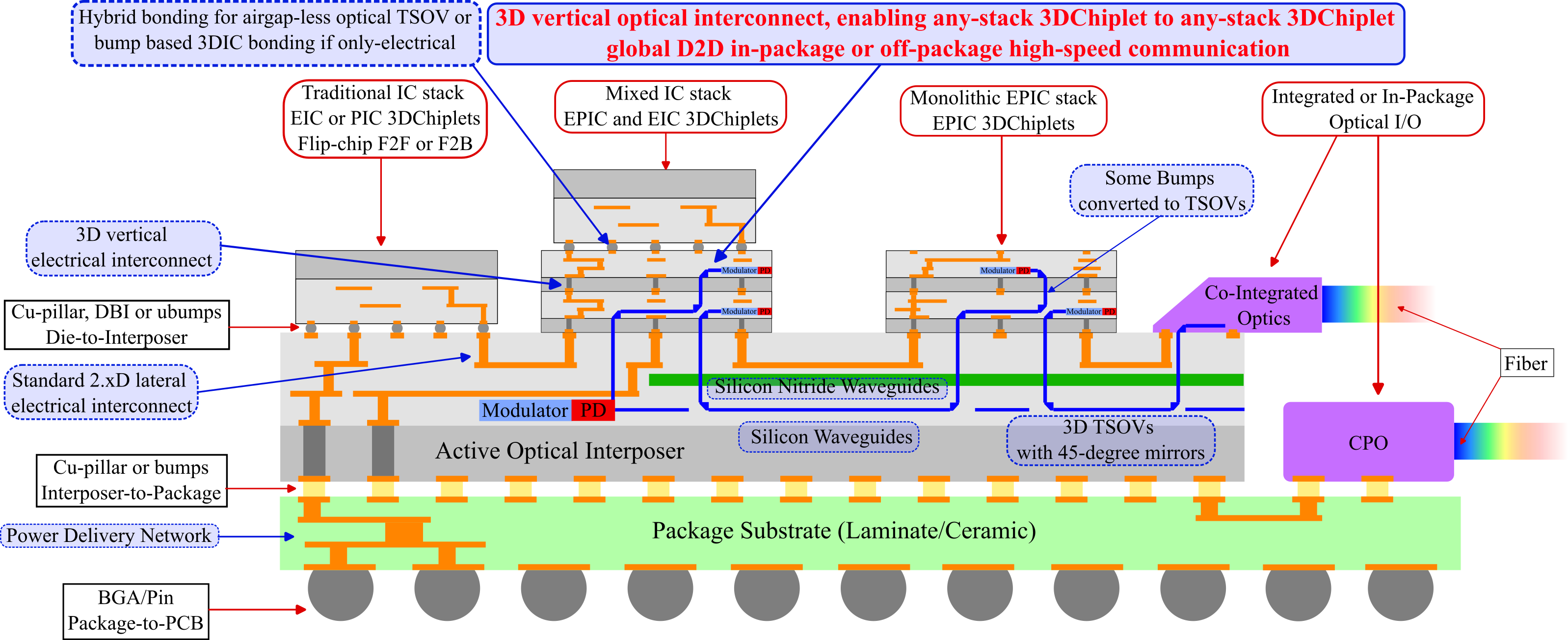}
		\caption{Proposed 3DChiplet-to-3DChiplet Electronic-Photonic Interconnect (3D-EPIC) platform. Scalable, energy-efficient, with massively parallel bandwidth throughput capabilities, with in-package optical interconnect and seamless optical connectivity to off-package optical interconnect.}
		\label{system_figure}		
	\end{figure*}
    Current chiplet architectures locate the Die-to-Die (D2D) communication chips very close, and move the package-level global links through a lower-speed data interface through the package substrate, or by employing an I/O switch chiplet in-package \cite{naffziger-AMD-2021}. These options are either channel-limited (for package links) or shoreline-limited. Copper interconnects combined with advanced packaging topologies continue to be the state-of-the-art solution for D2D applications. They employ 2.5D integrated multiple chiplets, flipchip bonded over a silicon interposer, which is then packaged on a build-up (BUP) laminate substrate \cite{cowos2023} or 3D integration where dies are 3D stacked over a base die, which is itself then integrated onto a substrate \cite{foveros2022,kim2024}. This strategy allows them to perform well for short links, but for longer links, the FoM drops by several orders of magnitude. Optical interconnect standards maintain a high FoM over very long distances due to low signal attenuation, or achieve a very high FoM over short distances due to energy efficiency and high bandwidth density. In Fig. \ref{fom_metric}, the OFC2022 datapoint is our system presented at \cite{anirban2023} and extended in \cite{chang2023}. UCD-TAMU EPIC datapoint is the maximum capacity of the same system before further scaling.
    
    Optical interconnect solutions, such as co-packaged optics (CPO), promise to overcome the challenges of off-package interconnect scalability; however, the electrical channel between the CPU/GPU/ASIC and the optical transceiver is a limitation, even within the package, when high-speed data has to traverse several tens of millimeters to reach the transceiver. A 3D heterogeneously integrated hybrid electronic-photonic interconnect platform, integrated over a silicon active optical interposer, can bridge this critical performance gap. Such a platform can enable a seamless, energy-efficient, high-bandwidth, fixed-low-latency data pathway between the compute cluster, the in-package memory, the I/O switch, and beyond. The 3DChiplet-stack electronic-photonic interconnect can connect every 3DChiplet in a stack to any other 3DChiplet in any other stack, thereby bringing all high-speed data transceiver nodes into direct access to the global interconnect. Enabling true scale-up or scale-out through resource disaggregation. Compatibility with high-performance, low-latency electrical interconnects remains critical in latency-sensitive applications such as cache-coherent memory. The optical interposer platform can support the highest density of 2.5D or 3D electrical interconnects, in addition to a photonic global interconnect. By leveraging novel technologies such as through silicon optical vias (TSOV), and through silicon vias (TSV) for the vertical interconnect and advanced silicon lithography for dense lateral wiring, such a platform will provide a unified platform for all applications. Supporting full backward compatibility for electrical-interconnect technologies such as HBM, and PCIe, while linking the CPO to a die-level optical interconnect. Heterogeneous integration allows electronic integrated circuits (EIC) and photonic integrated circuits (PIC) to be independently optimized, relieving the fabrication challenges by utilizing less dense technology nodes for the PICs (such as the monolithic electronic-photonic 45nm Fotonix\texttrademark$\ $ platform from GlobalFoundries \cite{gf45spclo}), and advanced process nodes for the CMOS EIC circuits. Resolving all the primary limitations for future AI HPC scalability.

	\section{Copper vs Photonic Die-to-Die (D2D) Interconnects}         
\begin{figure}[!t]
		\centering
		\includegraphics[width=3.4in]{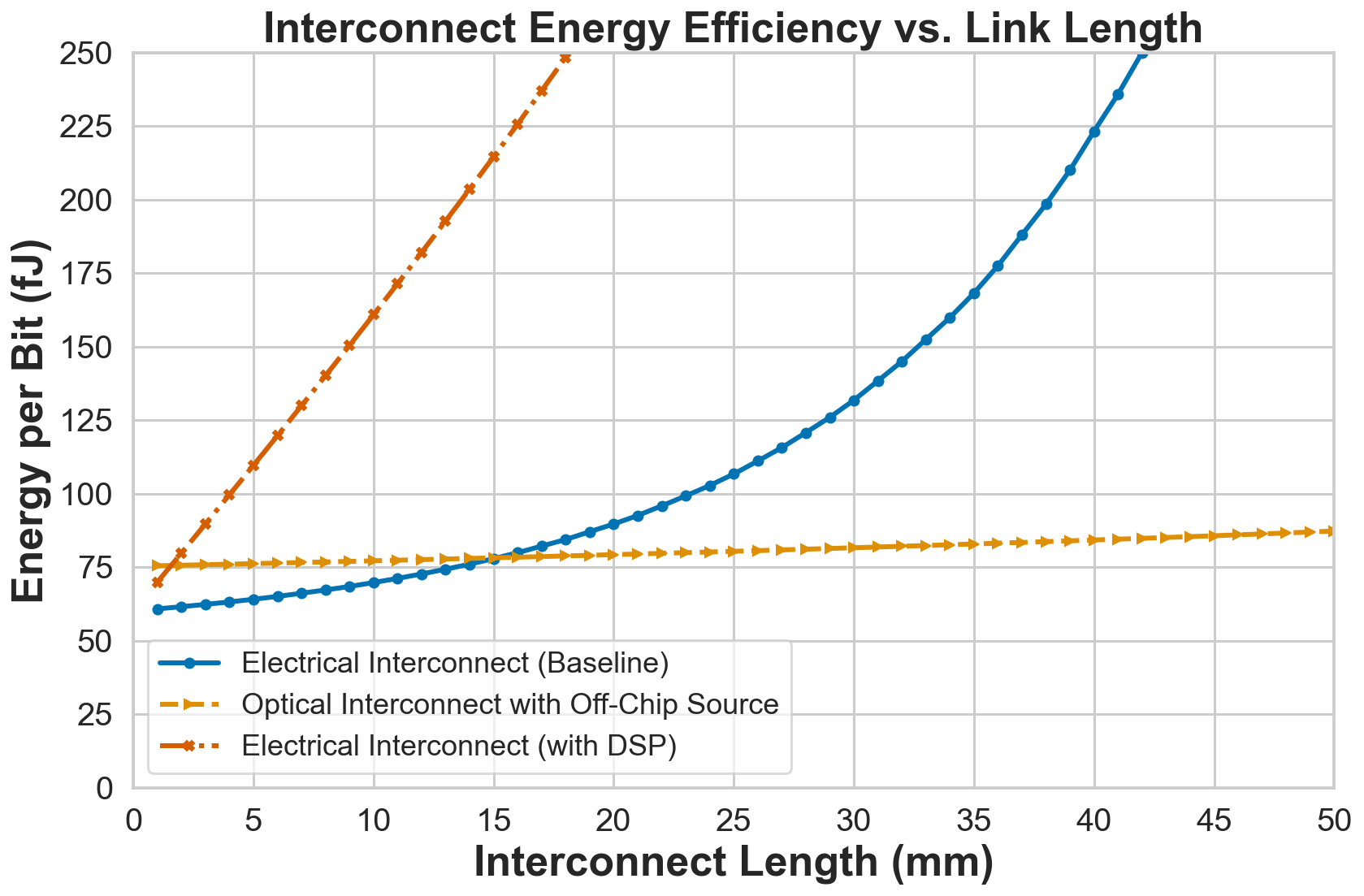}
		\caption{Comparing energy efficiencies for a copper interconnect and an optical interconnect at 8 Gbps datarate.}
		\label{interconnects_compared}
	\end{figure}

    For 3D photonics to replace state-of-the-art copper interconnects, it is essential to benchmark the various trade-offs in terms of the performance metrics discussed earlier. For 2.xD copper interconnects, the high-speed off-chip communication is implemented using transmission lines, with co-planar waveguide (CPW) models commonly adopted where signal integrity is critical. An electrical driver drives the signal into the CPW transmission line, and the signal propagates along the CPW with the electromagnetic fields guided in a quasi-TEM mode. At the link endpoint, the signal is received by an amplifier circuit and sampled. A simplified design space can be explored by varying the line width and spacing for a CPW model using the Wheeler approximation, while setting a target bandwidth and line rate for a fixed package-level link length and die edge length, which impacts the number of lines possible at the shoreline. For very thin line widths, increasing density becomes prohibitive due to attenuation loss for long links. For wider lines, the bandwidth density drops, failing to reach the target total bandwidth. In reality, the actual design space will be further limited due to signal impairments resulting from impedance mismatches in different transmission line designs, more complex crosstalk effects, and the inclusion of additional attenuation factors.
    \begin{table}[b!]
        \centering
        \caption{Parameters for Comparing Energy Efficiency vs Length}
        \label{tab:sim_params}
            \begin{tabular}{@{}lll@{}}
                \hline
                \textbf{Parameter Name} & \textbf{Value} & \textbf{Unit} \\
                \hline
                \multicolumn{3}{l}{\textit{\textbf{System Parameters}}} \\
                Bit Rate per Link & $8$ & Gbits/s \\
                Supply Voltage (VDD) & 1.0 & V \\
                \multicolumn{3}{l}{\textit{\textbf{Electrical Link Parameters}}} \\
                CPW T-Line Line Width & 2.0 & $\mu$m \\
                CPW T-Line Space to Ground & 2.0 & $\mu$m \\
                Relative Permittivity (Silicon interposer) ($\varepsilon_r$) & 3.9 & - \\
                Receiver Energy Cost & $60$ & fJ/bit \\
                \multicolumn{3}{l}{\textit{\textbf{Optical Link Parameters (Off-Chip Source)}}} \\
                Waveguide Loss & 1.0 & dB/cm \\
                Coupler Loss (per coupler) & 3.0 & dB \\
                Modulator Loss & 1.0 & dB \\
                Detector Responsivity & 1.0 & A/W \\
                Laser Wall-Plug Efficiency & 0.30 & - \\
                Modulator Capacitance ($C_\mathrm{MOD}$) & $50$ & fF \\
                Detector Load Capacitance ($C_\mathrm{LOAD}$) & $7$ & fF \\
                Modulator Driver Energy & $50$ & fJ/bit \\
                \hline
            \end{tabular}
    \end{table}
     To illustrate the energy efficiency of an isolated long electrical link vs an optical link, a simplified system simulation model is set up with the system parameters as tabulated in \ref{tab:sim_params}. The driver and receiver models are simplified to fixed energy costs. A range of interconnect lengths is simulated, and a voltage transfer function (VTF) method is used to determine the energy lost due to the transmission lines. The total energy cost due to the driver, interconnect loss, and the receiver constitutes the electrical link energy. The receiver requires a minimum amount of power to generate an output, and the light source must overcome the total loss in the system. Therefore, it's calculated backwards from the receiver sensitivity to the electrical power needed at the source. The energy efficiency is then calculated and plotted in Fig. \ref{interconnects_compared}. The crossover point, at which the optical interconnect outperforms the electrical interconnect, called the partition length \cite{Naeemi2004,beausoleil2008} occurs at $15.1\ mm$ in this model. The exponential growth of the energy per bit requirement of electrical interconnects, compared to the flat energy cost of optical interconnects, is well known \cite{stucchi2013}. In reality, even a 25mm long in-package electrical interconnect is considered extreme \cite{fan2024}; however, this can be made possible with photonic links. With modern GPU and CPU package sizes growing beyond 50mm edge length, edge-to-edge high-speed in-package communication can only be made possible with expensive I/O dies \cite{amdgenoa,naffziger-AMD-2021}. For future scalability, photonic links are the only option.
    This simulation simplifies the interconnect model and doesn't include any DSP components necessary to overcome signal integrity impairments, such as attenuation, reflections, crosstalk, and jitter, which require link optimizations with specialized digital signal processing (DSP) circuits, including equalization and error correction. These DSP circuits are very costly in terms of power, but are necessary in all but the shortest of D2D links. Depending on their complexity, they can encompass a wide range of energy costs. Table \ref{tab:dsp_cost} tabulates representative values from literature for state-of-the-art implementations. Although not a DSP component, clock and data recovery (CDR) circuits, in particular, tend to be very expensive due to their architectural complexity and strict jitter specifications, resulting in high dynamic power consumption \cite{kim2018}. Communication over optical interconnects can alleviate this issue by optically forwarding the clock signal from the source, thereby requiring a simpler and less energy-intensive CDR at the receiver.
    \begin{table}[h!]
        \centering
        \caption{Energy cost of High-Speed Link Components}
        \label{tab:dsp_cost}
        \begin{tabular}{@{}llc@{}}
            \hline
            \textbf{Function} & \textbf{Energy Efficiency} & \textbf{Ref.} \\
            \hline
            FEC & 20 fJ/bit & \cite{netherton2024} \\
            DFE & 27 fJ/bit/dB & \cite{chen2014} \\
            CTLE & 50 fJ/bit & \cite{manian2017} \\
            CDR & 1900 fJ/bit & \cite{kim2018} \\
            \hline
        \end{tabular}
    \end{table}
    The signal attenuation in dB/cm in optical interconnects is at least an order of magnitude lower compared to copper, 1 dB/cm for Si waveguides, and 0.024 dB/cm for SiN waveguides \cite{liu2021} compared to 3-10 dB/cm for typical narrow copper interconnects on silicon interposers. Due to this, optically interconnected systems can implement simpler circuits rather than the expensive DSP circuits, leading to better link energy efficiency, as shown in Fig. \ref{interconnects_compared}. The energy cost of adding DSP circuits would eliminate the partition length, making optical interconnects superior in terms of energy efficiency, even for short links. Adding a channel loss-dependent DSP cost (such as DFE) to our model reduces the partition length to 2.5mm. The requirement of complex circuit elements, such as equalization for long electrical interconnects, is a critical pain point in high-speed communication. This cost rises sharply as the link length increases, due to signal impairments that need to be mitigated. In the proposed 3DChiplet-stack electronic-photonic interconnect platform, the electrical link over TSVs maintains the current advantages of 3D-IC architectures, i.e., reducing the link length in lossy interposer dielectrics to the order of die thickness or approximately the TSV height, resulting in an order-of-magnitude improvement. It also maintains the ability to have a short-reach 2.5D interconnect to a lateral adjacent die, such as current HBM implementations. This is desirable for latency limits in cache-coherent applications. Although the electrical channel lengths in 3D-IC packages are short, reaching a global interconnect, either in-package interface or system interconnect standards such as PCIe with copper still require long link lengths and remain a bottleneck that 3D photonics can resolve.
    
    Link latency is a critical part of any communication link. Although the lowest possible latency is always desired, there are trade-offs with complexity, cost, and power consumption. Signal integrity effects, such as crosstalk-induced delay, which can result in latency, are mitigated by transmission line design or through the use of complex SerDes circuits. Optical interconnects are not affected by the same latency challenges. However, the need to convert between electrical and optical domains at the link endpoints adds latency to any optical link. This makes electrical links still desirable in some very short-link, dense I/O applications where the minimum possible latency is desired at the cost of energy efficiency. For most other applications, optical links for D2D and system-level links are superior. The proposed 3DChiplet-stack electronic-photonic interconnect platform maintains this ability for 2.5D heterogeneous integration of XSR-linked ICs, while bringing the global low-latency optical interconnect to every high-speed data transmission node.

    Bandwidth density is dependent on the physical topology of the die-package. In current 3D-IC stacks, the bandwidth density is a function of the areal bump density, as each signal pin has to be assigned to bumps for off-die interconnect. Here, we compare our proposed interconnect with the UCIe-3D interconnect, which is the current state-of-the-art 3D electrical interconnect standard. Das Sharma et al. outline a process for benchmarking the bandwidth density of advanced interconnects in \cite{ucie2024} for the UCIe-3D standard and in \cite{ucie2022} for the published UCIe-2D standard.
    \begin{equation}
        \label{bw_density_theoratical}
        \text{$BW\ Density_{\ Theoratical}$} =\\ 
        \frac{Datarate_{\ channel}}{\text{${Bump}^2_{pitch}$}}
    \end{equation}
    However, \cite{ucie2024} states that the actual realizable bandwidth is limited by practical overheads, as some bumps need to be assigned to power/ground and repair lanes. A bump efficiency factor is required to account for the different physical bump patterns of square and hexagonal arrays.
    \begin{equation} \label{bw_density_realizable}
        \begin{split}
        \text{$BW Density_{\ Realizable}$} &=
          \text{$BW\ Density_{\ Theoratical}$} \\
                          &\quad \times\ \eta_{\text{bump}} \times \bigl(1 - OH_{\text{total}}\bigr)
        \end{split}
    \end{equation}
    where $\eta_{\ bump}$ is the bump efficiency factor, set to 1.15 for hexagonal arrays and 1 for square arrays. Hexagonal arrays are common in less dense interconnects, such as traditional copper-pillar (CuP) implementations. In contrast, denser implementations, like Direct Bond Interconnect (DBI\textsuperscript\textregistered), utilize a square pattern due to fabrication and assembly challenges. The $OH_{\ total}$ is the overhead factor. This includes $\sim$3$\%$ for the data and $\sim$10$\%$ for repair overheads in dense 3D interconnects or $\sim$3$\%$ for 2.xD interconnects, we here assume at 25um pitch and above, as less dense interconnects are less prone to failure \cite{ucie2024}. Power and ground overheads are defined in parts as a function of bump pitch, which is dictated by signal and power integrity considerations. This is a critical design factor in electrical interconnects for 3D-IC packaging and is not a linear requirement due to trade-offs in signal and power integrity. At higher bump pitches, the channel datarate can be higher with greater spacing between signal bumps, but requires additional ground-shielding bumps for signal integrity. At lower pitches, the bump and the associated TSVs are much narrower and can support lower current capacity, requiring more bumps to maintain power integrity in the power delivery network (PDN). The maximum data rates are also defined by parts for different bump pitch ranges; therefore, they are set according to the UCIe-3D specifications. 

    For the 3D optical bandwidth density in our proposed platform, we envision converting a fraction of the bumps across the 3DChiplet area into TSOVs. We define this as a TSOV conversion ratio ($R_{\ TSOV} = {TSOV/\ bump}$), for example, $R_{\ TSOV} = 2\%$ indicates 1 in 50 signal bumps is converted to a TSOV. We first estimate the bump density across the 3DChiplet area ($D_{bumps}$, from the square-grid packing estimate of a lattice defined by the $Bump_{pitch}$) and then use the conversion ratio to determine the TSOV density. This is a reasonable estimate given that the TSOV structure that we are developing is $<1\mu m$ in areal footprint.
    \begin{equation}
        \label{num_of_tsov}
        \text{$D_{TSOV}$} =\\ 
        \text{$D_{bumps}$} \times R_{TSOV}
    \end{equation}
    \begin{equation}
        \label{bw_optical}
        \text{$BW Density_{\ optical}$} =\\ 
        \text{$D_{TSOV}$} \times N_{WDM} \times Datarate_{channel}
    \end{equation}
    \begin{figure}[b!]
		\centering
		\includegraphics[width=3.4in]{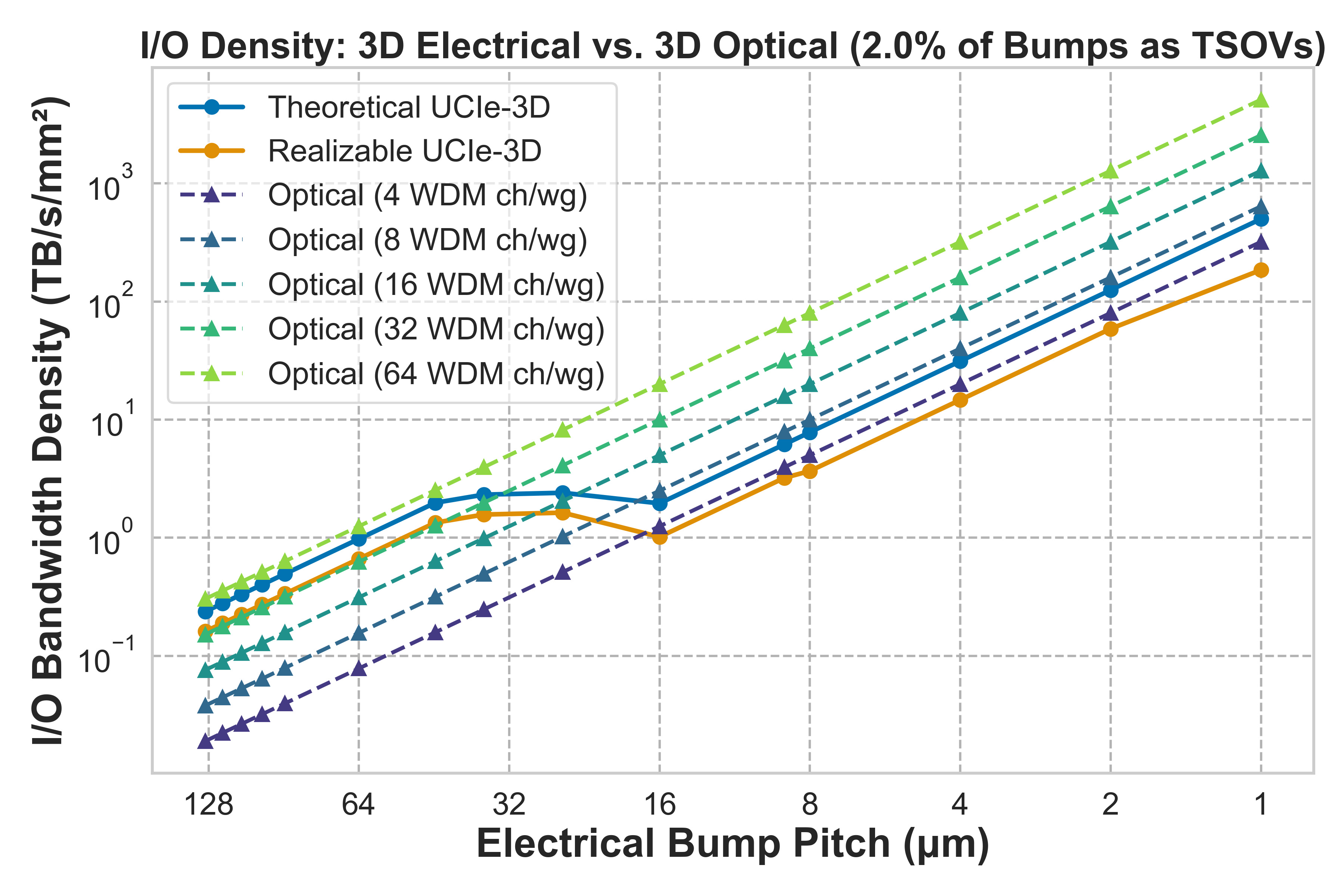}
		\caption{The total bandwidth density of a theoratical 3D optical interconnect vs the projected UCIe-3D standard. The optical interconnect here assumes a fixed channel datarate, and WDM channel count per TSOV.}
		\label{bw_density_compared}
	\end{figure}
    where $N_{WDM}$ is the number of WDM channels per TSOV. Fig. \ref{bw_density_compared} plots the $BW Density_{\ optical}$ along with the theoretical and realizable bandwidth density of UCIe-3D at different bump pitches. We don't curve fit the UCIe-3D datapoints here, and the x-axis is flipped to show the increasing bandwidth density with denser interconnects. The optical interconnect assumes that there is a fixed number of multiplexed wavelengths per TSOV channel, and a chiplet of 1mm x 1mm is considered featuring a bump pitch of 55$\mu$m. This is chosen as it is the standard pitch of HBM 3D-stack modules for a possible real application scenario. Several variations of the WDM channel count are plotted, from 4 channels to 64 channels per waveguide/TSOV. The 3D electrical interconnect notably decreases in bandwidth density around the 32$\mu$m bump pitch. This is due to the piecewise definition of the power/ground overheads and channel data rates. This corresponds to the different IC packaging paradigms with electrical interconnects. The higher bump pitch corresponds to 2D packaging, the range around 32$\mu$m pitch corresponds to 2.5D packaging, while the denser pitches correspond to 3D packaging. As an example, with the stated parameters, the chiplet has 330.6 bumps/$mm^2$, rounded to 330 and assumed to be in a hex pattern, with an overhead of 0.39, resulting in an electrical bandwidth density of $925.98\ GB/s/mm^{2}$. The corresponding calculations for the 3D optical interconnect with 32 channels get a bandwidth density of $768\ GB/s/mm^{2}$. Although the 3D electrical interconnect outperforms in this regard, the advantage of optical communication lies in its scalability with WDM. By increasing the multiplexed wavelength channels per TSOV to 39, the optical interconnect bandwidth density increases to $936\ GB/s/mm^{2}$ and outperforms the electrical interconnect. Large WDM systems require spectrum planning, while not trivial it's beyond the scope of the current benchmark study of potential bandwidth density. At denser bump pitches, even a simpler 4-channel WDM optical interconnect outperforms the electrical interconnect in terms of bandwidth density at a given TSOV conversion ratio as the channel datarates are not constrained due to signal integrity.  3DChiplet optical interconnects excel in this regard due to their ability to accommodate bandwidth density requirements across every pitch range within a standard design philosophy. This would reduce the complexity of the design across a range of 3D-IC applications. This demonstrates that our 3D EPIC platform can beat the 3D electrical interconnects to $>$10\ TB/s/$mm^2$ bandwidth density and beyond.
    \begin{figure}
		\centering
		\includegraphics[width=3.4in]{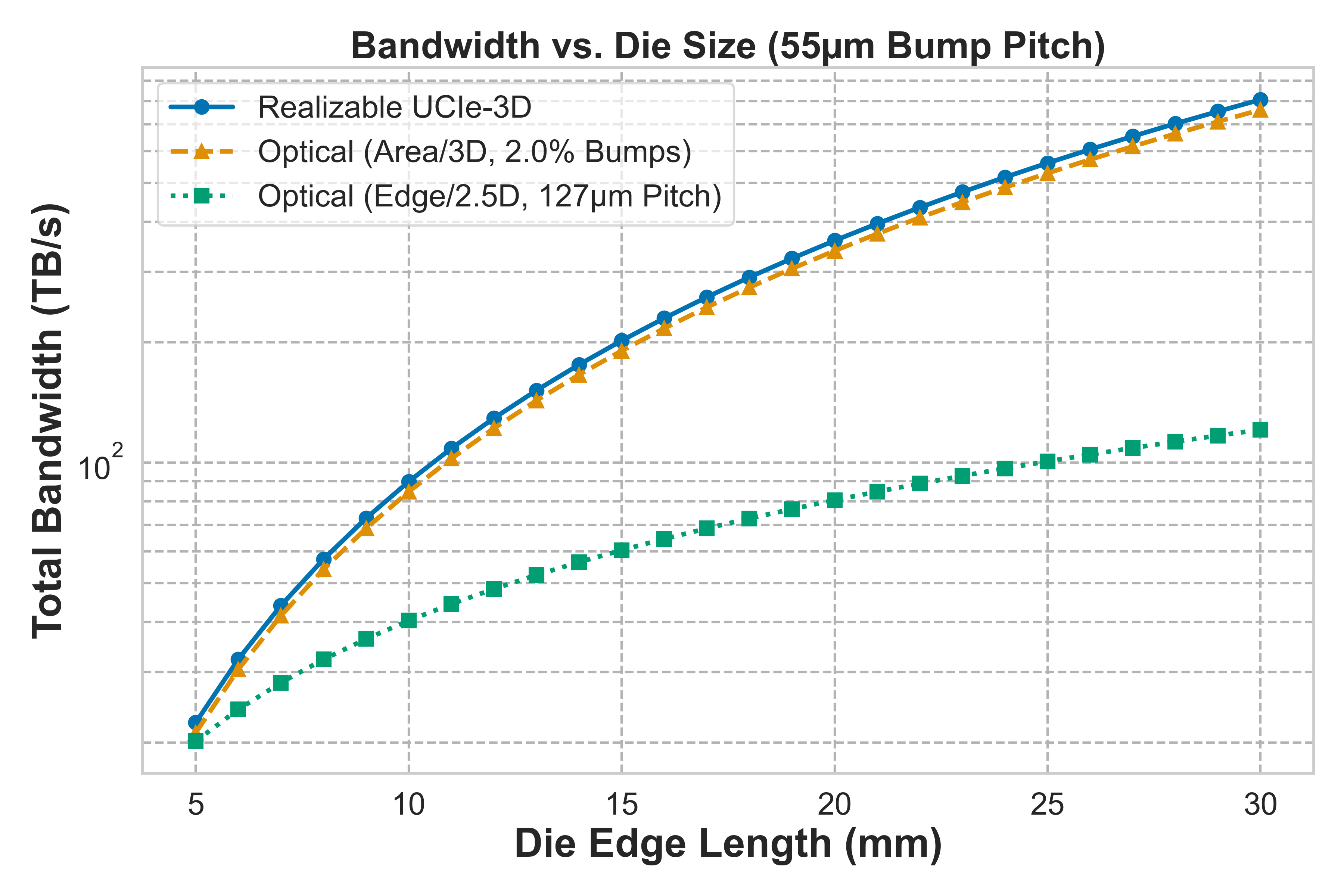}
		\caption{Total achievable bandwidth of a theoratical 3D optical interconnect compared against the projected UCIe-3D standard plotted vs the die size set by the die edge length. The optical interconnect here assumes a fixed number of 32 channels at 32 Gbps datarate per channel, and the electrical bump pitch is fixed to 55$\mu$m.}
		\label{bw_total}
	\end{figure}

    Fig. \ref{bw_total} plots the total bandwidth for a 55$\mu$m bump pitch design and varies the die edge length, and hence the die size. For 3D interconnects, this means the product of the bandwidth density with the chiplet area, as the entire area is utilized to define the interconnect. At 55$\mu$m pitch, just a 2$\%$ TSOV bump conversion rate is sufficient to match the 3D electrical interconnect, and Fig. \ref{bw_density_compared} shows that 3D optical outperforms at denser pitches. High-speed signal transmission in copper at extremely low pitch ranges is challenging due to crosstalk; in this context, 3D photonics outperforms electrical interconnects. Optical signals in silicon photonic waveguides are exceptionally well confined, with the evanescent field becoming negligible at less than 100nm beyond the boundary. If all high-speed signal transmission is moved to the 3D optical interconnect with the electrical interconnect handling the PDN, then such extremely dense interconnects could be made possible, making a 3D optical interconnect necessary. The figure also includes a 2.5D optical interconnect scenario. For this scenario, the perimeter of the die is calculated from the die edge, and dividing by a standard fiber array pitch of 127$\mu$m gives the number of fiber I/O. The data rate and number of WDM channels per waveguide remain consistent with the number of WDM channels per TSOV. This illustrates the shoreline limitation of 2.xD interconnects. Area-based interconnects scale with the square of the die edge ($\propto L^2$), while those based on die perimeter or shoreline scale linearly ($\propto L$). As chips become larger, area-based 3D I/O interconnects gain a significant advantage in terms of bandwidth and bandwidth density metrics.

    \begin{figure}[h!]
		\centering
		\includegraphics[width=3.4in]{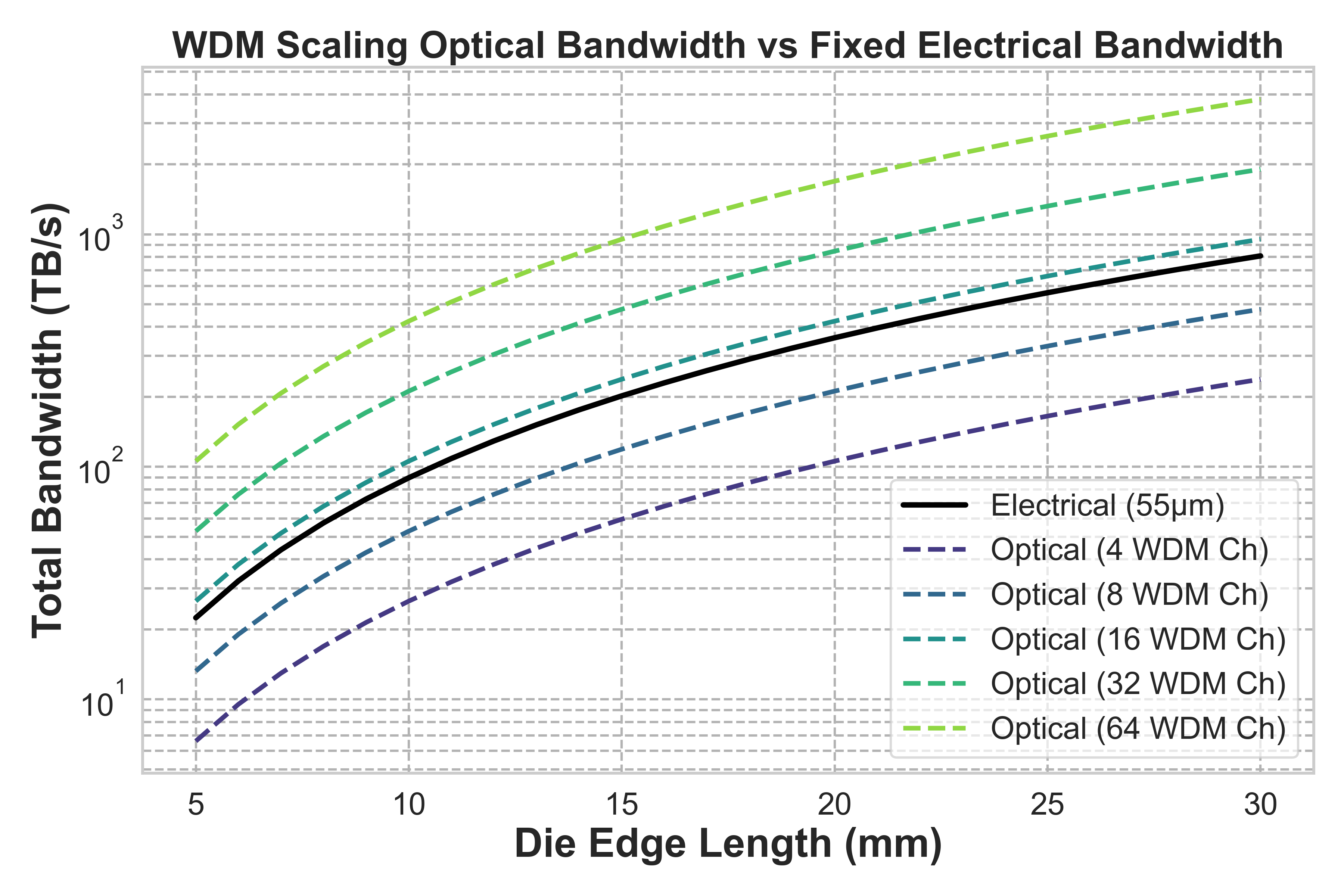}
		\caption{The effect of WDM on the total achievable bandwidth of a theoretical 3D optical interconnect vs the projected UCIe-3D standard. The number of multiplexed transceiver channels is varied from 4 to 32. The TSOV-bump conversion is set to 5$\%$, and channel data rate at 32 Gbps.}
		\label{link_wdm_fixed_electrical_ref}
	\end{figure}
    \begin{figure}
		\centering
		\includegraphics[width=3.4in]{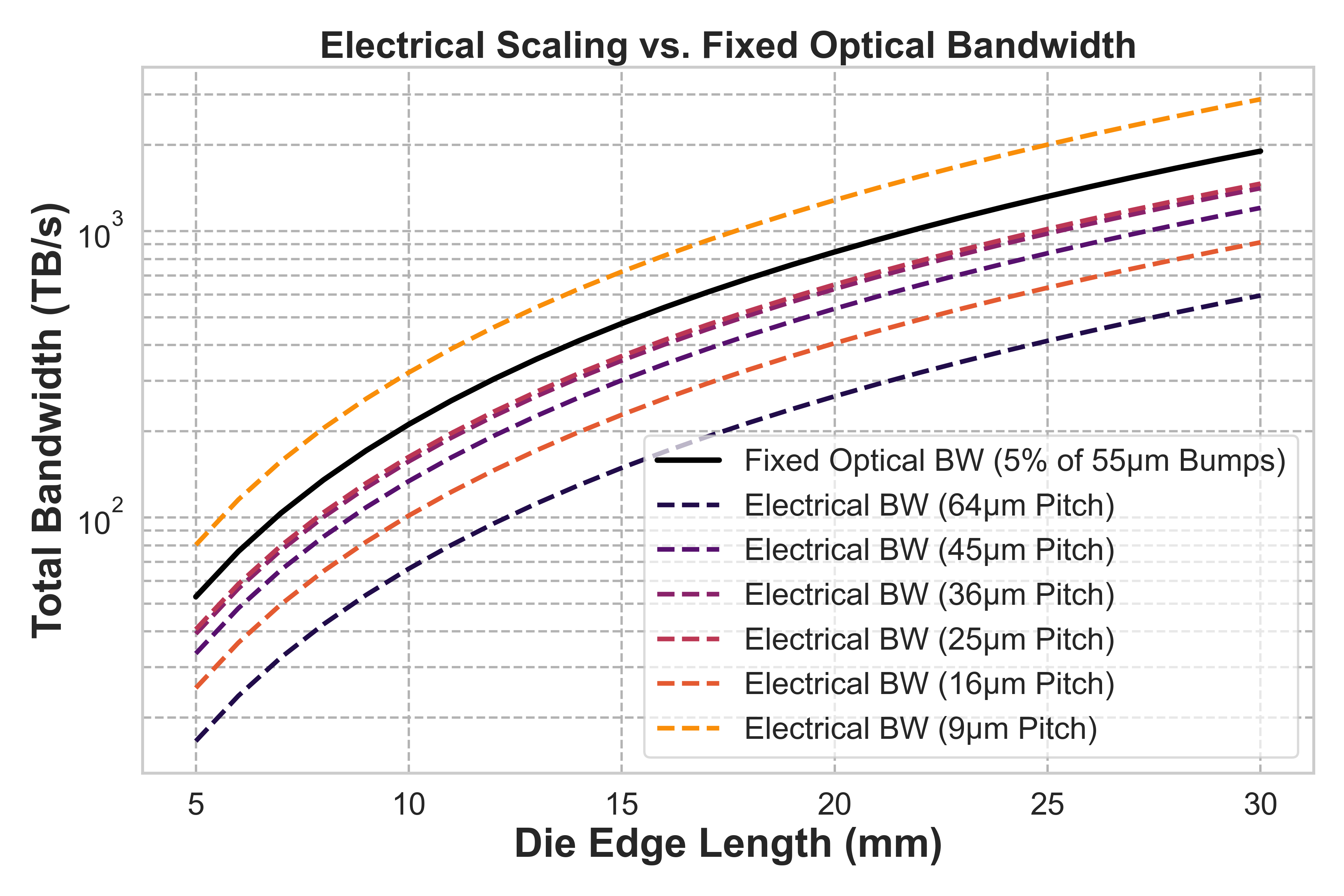}
		\caption{Alternative interconnect decision space when considering how dense an electrical-only interconnect needs to be to beat the bandwidth throughput of a 3D optical interconnect.}
		\label{link_electrical_fixed_optical_ref}
	\end{figure}
    
    Optically interconnected high-speed communication has a significant advantage in scalability due to WDM. When planning for the interconnect, decision-making must consider the design space.  Fig. \ref{link_wdm_fixed_electrical_ref} shows how changing the count of wavelength-multiplexed channels per TSOV affects the total achievable bandwidth, assuming a fixed number of TSOVs. A $5\%$ TSOV conversion rate is fixed to replicate a fabrication-limited count of TSOVs, and the bump density is determined by the bump pitch of 55$\mu$m. An alternative approach to examining this problem is to determine the density of a 3D electrical interconnect required to match a 3D optical interconnect. Fig. \ref{link_electrical_fixed_optical_ref} plots various total bandwidth outcomes of the electrical interconnect bump pitches against a reference 3D optical interconnect. We set a 5$\%$ TSOV conversion rate at a 55$\mu$m bump pitch for the optical interconnect as a baseline. The WDM wavelength count is kept at 32. Only the highest density electrical can outperform the optical interconnect. Instead of going to a 9$\mu$m pitch electrical interconnect and taking on the complexity of signal integrity and PDN for such a dense system, the same bandwidth can be achieved by using the 3D optical interconnect while retaining a 55$\mu$m pitch electrical interconnect for power delivery.    
    Comparing the performance metrics, the proposed 3D-EPIC platform addresses the primary limitations of interconnects and can be transformative for high-performance computing applications, enabling future scalability for energy-efficient, high-bandwidth, high-speed D2D communication links.
    
	\section{3D-Chiplet Stacked Electronic-Photonic Interconnect Platform}
    We are simultaneously developing several technologies necessary to enable the 3D-EPIC platform, and this section covers our progress. The through-silicon optical via (TSOV) for the vertical transition of the optical link is the most critical feature of the platform, enabling the vertical optical signal transition between 3D chiplets and linking the 3D chiplet stack to the optical signal plane in the active optical interposer. The concept was first proposed by Zhang et al. in \cite{zhang2018} as a vertical u-bend between multiple photonic waveguide layers within the same PIC. With progress in the fabrication process since then, the concept can be extended in the Z-axis to inter-chiplet TSOVs. Like traditional TSVs, except that the vertical via being optimized for low-loss optical transmission.
    \begin{figure}
        \centering
        \includegraphics[width=3.4in]{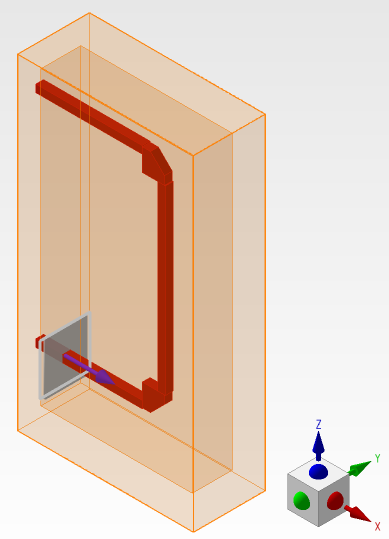}
        \caption{Ansys Lumerical FDTD model of TSOV for simulation. Includes two lateral waveguide sections, two 45\textdegree mirror sections and a vertical waveguide section, which serves as the TSOV. Simulation shows less than 1dB loss per TSOV.}
        \label{tsov_model}
    \end{figure}
    
    The TSOV marks a substantial development in vertical bandwidth density compared to TSVs. It has several advantages over traditional vertically launched optical coupling methods, such as wavelength-dependent grating and evanescent coupling, in a drastically reduced footprint. It consists of a 45\textdegree mirror and a vertical waveguide section. Two lateral waveguide sections on the top and bottom die complete the via structure. The structure was modeled and simulated in Lumerical FDTD. Swarm optimization was used to determine the optimal and worst-case coupling loss between the mirror and the waveguide segments, aiming to align them. With no offsets, a coupling loss of 0.34 dB was observed; however, with precise nanometer-scale offset engineering, a coupling loss of 0.24 dB is achievable. For the no-offset scenario, a coupling loss of 0.7 dB was achieved for the whole TSOV, with a best-case loss of 0.42 dB possible with offsets. Across the simulated C-band spectrum, the coupling loss remained below 1 dB. Bosch etching was used to etch the high-aspect ratio via in crystalline silicon. We have currently demonstrated fabrication of 20$\mu$m high via structure, which is 54:1 aspect ratio for a width of 370nm. We aim to develop up to 100$\mu$m high TSOVs for an aspect ratio of 270:1. The etching process creates a corrugated surface, however current simulation shows that this does not increase the loss for the optimized TSOV model. We are still developing the fabrication process; however, the reported results demonstrate that TSOVs can be reliably fabricated in a CMOS-compatible silicon process.
    \begin{figure}
		\centering
		\includegraphics[width=3.4in]{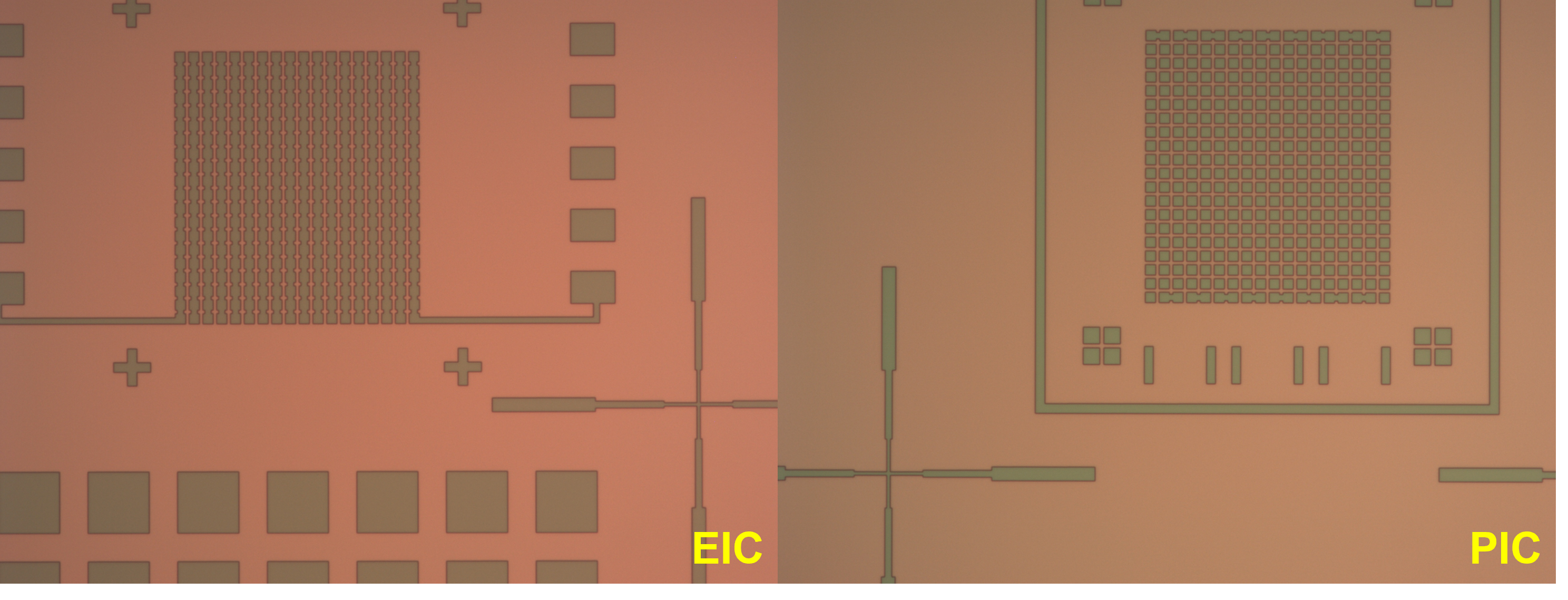}
		\caption{Dummy EIC and PIC matching designs with daisy chain test structures to characterize connectivity.}
		\label{eutectic_die_test}
	\end{figure}
    \begin{figure}
		\centering
		\includegraphics[width=3.4in]{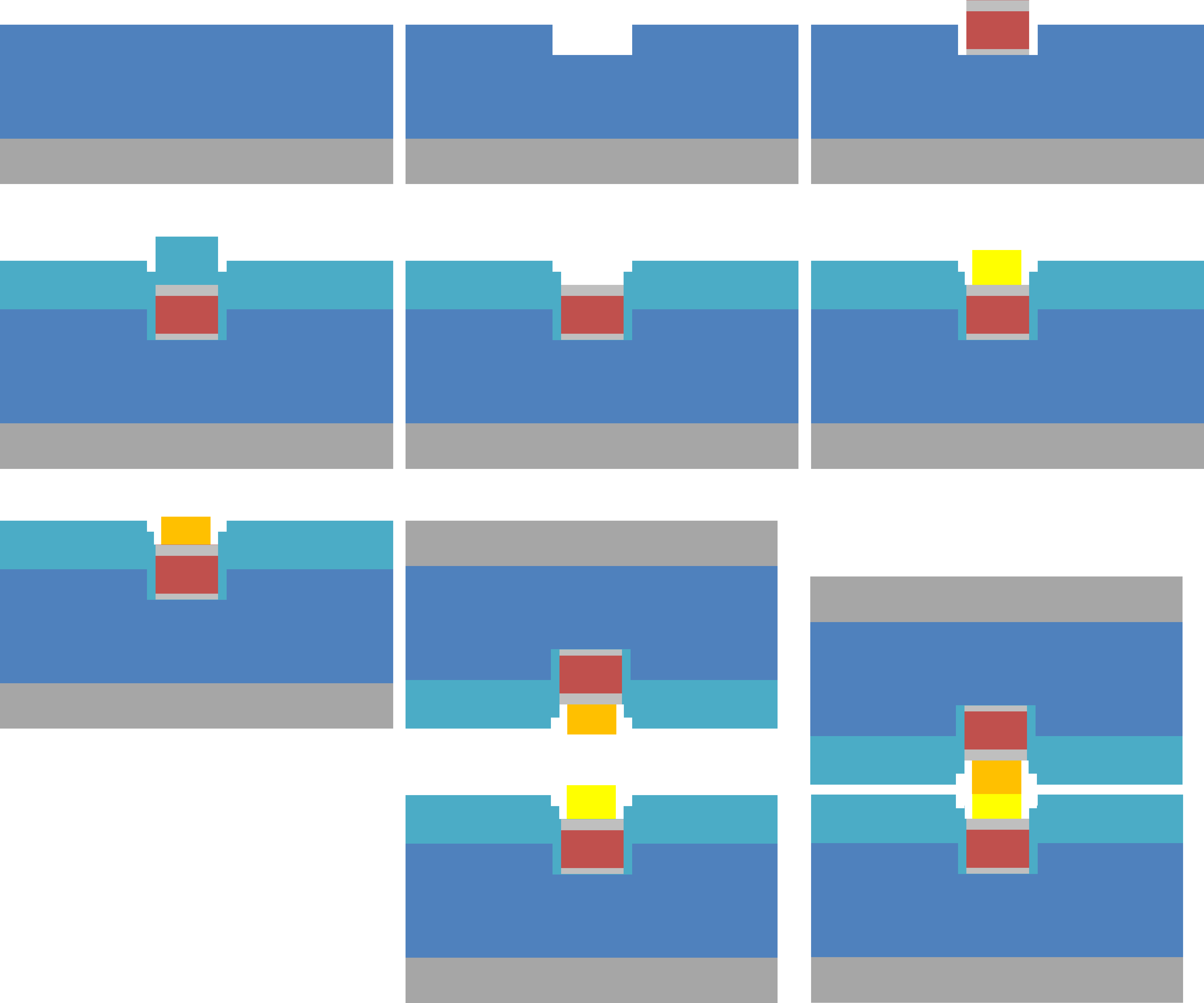}
		\caption{AuSn20 eutectic metallurgy D2D bonding process development. Continued bonding process development will enable an airgap-less D2D bonding solution, eliminating the risk of solder reflow associated with traditional solder bump bonding.}
		\label{eutectic_process}
	\end{figure}
    
    For a vertical TSOV-enabled 3D photonic interconnect, we anticipate that an airgap-less 3D integration process, such as hybrid bonding or DBI\textsuperscript{\textregistered}, will ideally be required. Currently, DBI\textsuperscript{\textregistered} in it's most advanced node is still only a die-to-wafer (D2W) bonding technology \cite{dbi2020}, with plans to adapt it to D2D bonding. We've been pursuing D2D bonding as part of our interposer bonding solutions. Although we would like to explore hybrid bonding eventually, at this time, we are developing 80-20$\%$ by weight Gold-Tin eutectic (AuSn20) metallurgy as a compromise between solder bump based bonding and hybrid bonding. As an advantage over solder bump bonding, this method allows for multi-stage assembly without the risk of reflow. Our current bonding process is shown in Fig. \ref{eutectic_process}. We first create dummy test dies, shown in Fig. \ref{eutectic_die_test} on silicon wafers by growing wet oxide on the wafers followed by oxide etching, and sputtering Nickel-Copper-Nickel to form test bond pads. We use a lift-off process at this stage of development. The first thin layer ($\leq$20nm) of nickel is used for adhesion on oxide, followed by copper as the pad metal, and then another thin layer (30nm) of nickel for surface protection. The sputtering is performed in-situ in vaccuum to prevent any surface contamination. We also form alignment markers, and lithography characterization markers at this stage. Next we deposit passivating oxide with Plasma-Enhanced Chemical Vapor Deposition (PECVD) at 300\textdegree C, and $< 5\times10^{-5}$ Torr pressure. Next we oxide etch to clear the bond pad surface, followed by second lithography step to form bumps. The etch window here is larger than the pad, forming a cavity well for the bump material to reflow and fill. The nickel layer cap prevents the top surface from oxidizing between process steps and enhances the adhesion of subsequent layers. To form the bumps we deposit gold on one wafer, and thermally evaporate AuSn20 eutectic compound on the other mating wafer. This is for the experimental stage due to current cleanroom limitations. In future, all bumps will be the eutectic material. Following a lift-off process, the wafers are diced to singulate the test dies. We intentionally form tall bumps during alignment characterization, thus leaving an airgap after bonding. Eventually, the bump volume will be reduced to fill the bump cavity exactly and eliminate the air gap. A Ficontec FL300 die-bonding tool is then used to flip-chip bond the two dies. Alignment testing samples are fabricated on double-side polished wafers to characterize bonding alignment with an IR scope. Probe tests were conducted to verify connectivity. Although at low yield, we successfully demonstrated connectivity across a 2500 pad daisy-chain test structure. Finally, a shear force tester is used to characterize the bond strength, with our best sample withstanding a shear force of 22 Kg-N force. The die sample fractured without the bond breaking in this case. The thermally evaporated AuSn20 deposition step should ideally be performed in a forming gas environment, which is not possible in our cleanroom. We discovered through Energy Dispersive Spectroscopy (EDS) x-ray analysis that the deposited material composition had shifted towards more gold, resulting in void formation in the bump metal. However, starting from an off-eutectic compound of AuSn24 resulted in AuSn20 compound after thermal evaporation, allowing for successful bonding.
    \begin{figure*}
		\centering
		\includegraphics[width=6.8in]{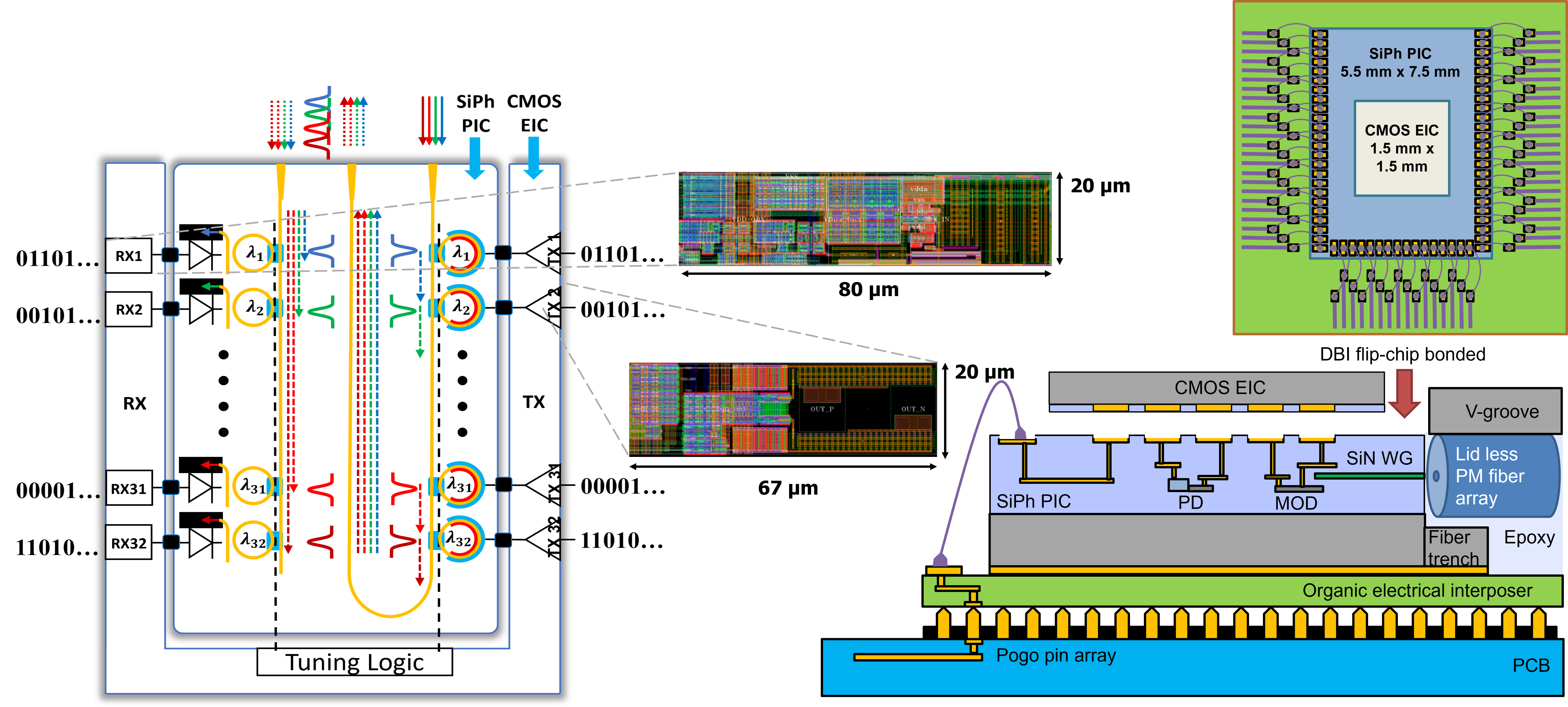}
		\caption{System architecture of a 3D heterogeneously integrated electronic-photonic high-speed transceiver module. Previously reported in \cite{chang2023,anirban2023}, this co-designed EIC-PIC architecture demonstrated record low receiver sensitivity at the time by utilizing the lowest parasitic 3D integration DBI\textsuperscript{\textregistered} process.}
		\label{doe_architecture}
	\end{figure*}
    \begin{figure}
		\centering
		\includegraphics[width=3.4in]{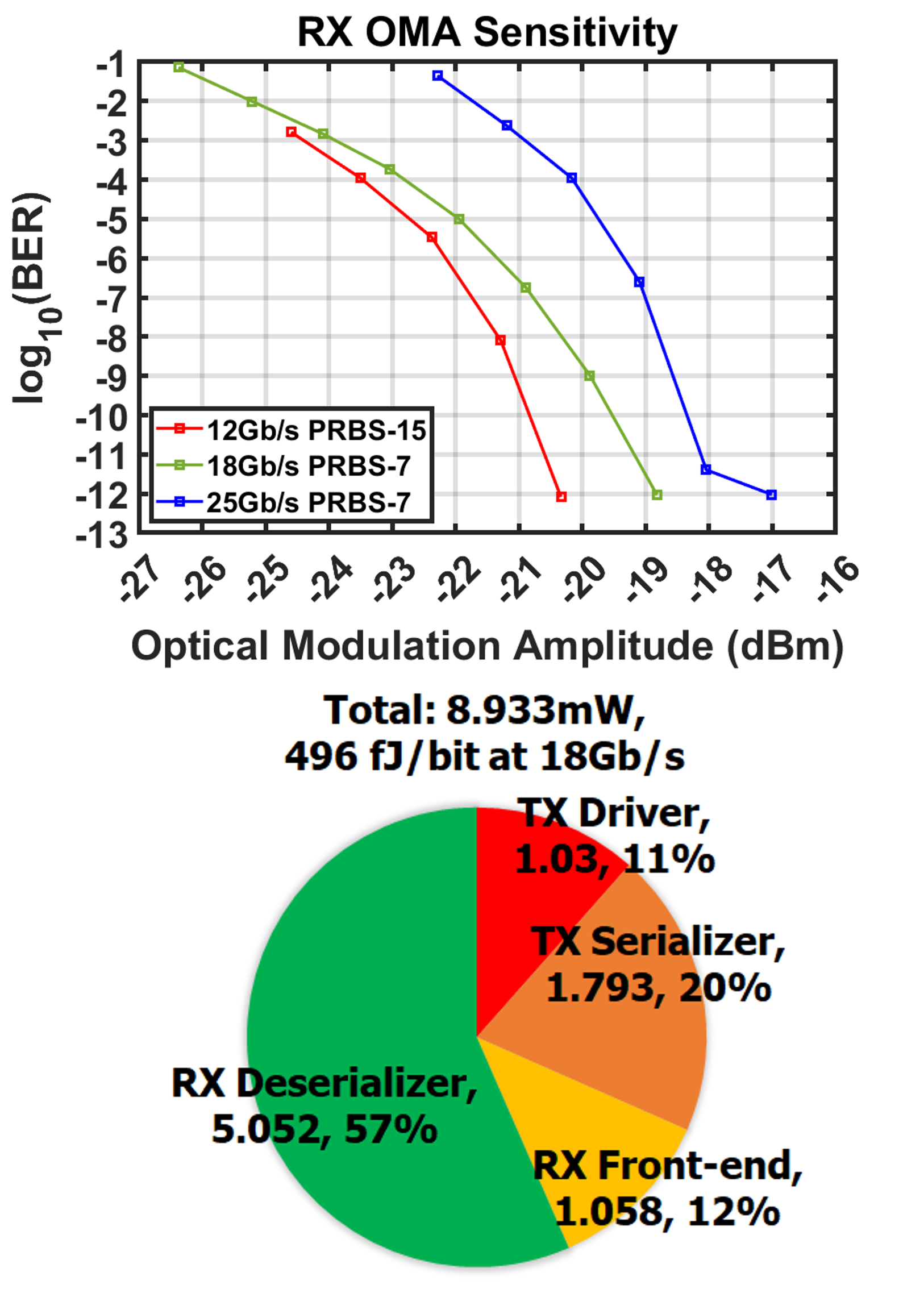}
		\caption{Receiver OMA sensitivity and energy efficiency breakdown of the SerDes reported in \cite{chang2023}. At the time of reporting, -17.01 dBm at 25 Gb/s was the lowest receiver OMA sensitivity. The EIC SerDes achieved 496 fJ/bit in a 12nm process node, and we expect process contraction to 3nm to enable a pathway to transceivers with a power consumption of $\leq 100\ fJ/bit$.}
		\label{doe_oma}
	\end{figure}
    
    \begin{table}[b!]
        \centering
        \caption{Extracted Parasitics using Ansys Q3D \\ @ 5GHz}
        \label{tab:pkg_parasitics}
        \begin{tabular}{@{}llll@{}}
            \hline
            \textbf{Bump type} & \textbf{Pad Size-Pitch} & \textbf{Resistance (m$\Omega$)} & \textbf{Capacitance (fF)}\\
            \hline
            DBI\textsuperscript{\textregistered} & 10$\mu$m-20$\mu$m & 19.36 & 7.2\\
            C4 & 80$\mu$m-160$\mu$m & 13.71 & 31.6\\
            \hline
        \end{tabular}
    \end{table}
    The next critical enabler for high-speed optically interconnected communication links is a low-parasitic electrical interconnect between the PIC circuits and the serializer-deserializer (SerDes) control electronics. Monolithic processes feature the lowest possible parasitics; however, the state-of-the-art monolithic electronic-photonic fabrication nodes (currently at 45nm) are several generations behind the state-of-the-art CMOS process node. Heterogeneous integration enables EIC circuits to be designed, optimized, and fabricated in more advanced nodes, while PIC circuits are fabricated using a monolithic electronic-photonic process. In a heterogeneous EIC-PIC system, where the photodetector is located in the PIC and the TIA is located in the EIC, the optical link efficiency is determined by the minimum detectable current at the TIA. The additional parasitic impedance induced by packaging makes it a critical design component in system planning. We simulated a relaxed DBI\textsuperscript{\textregistered} bump design and compared to traditional C4 bumps. The simulated values are tabulated in \ref{tab:pkg_parasitics}. With the total interconnect input capacitance being the critical component to TIA performance, and even our relaxed DBI\textsuperscript{\textregistered} bump showing nearly 1/5 of the C4 bump capacitance, we decided to adopt DBI\textsuperscript{\textregistered} for 3D integration of our EIC and PIC. In \cite{chang2023}, we used this strategy to demonstrate the lowest reported receiver OMA sensitivity at the time of -17.01 dBm at 25 Gb/s, and -18.82 dBm at 18 Gb/s with NRZ signaling. This resulted in a SerDes energy efficiency of 496 fJ/bit at 18 Gb/s data rate. Although the transmitter was limited to 18 Gb/s due to foundry constraints, the receiver was independently tested at up to 25 Gb/s, resulting in an energy efficiency of 191 fJ/bit. The drastic reduction in interconnect parasitic capacitance from the DBI\textsuperscript{\textregistered} process was the key enabler for our receiver performance. Developed in partnership with industry partners, our 3D EPIC transceiver solution is industry-ready through technology transfer. 

\begin{figure}
		\centering
		\includegraphics[width=3.4in]{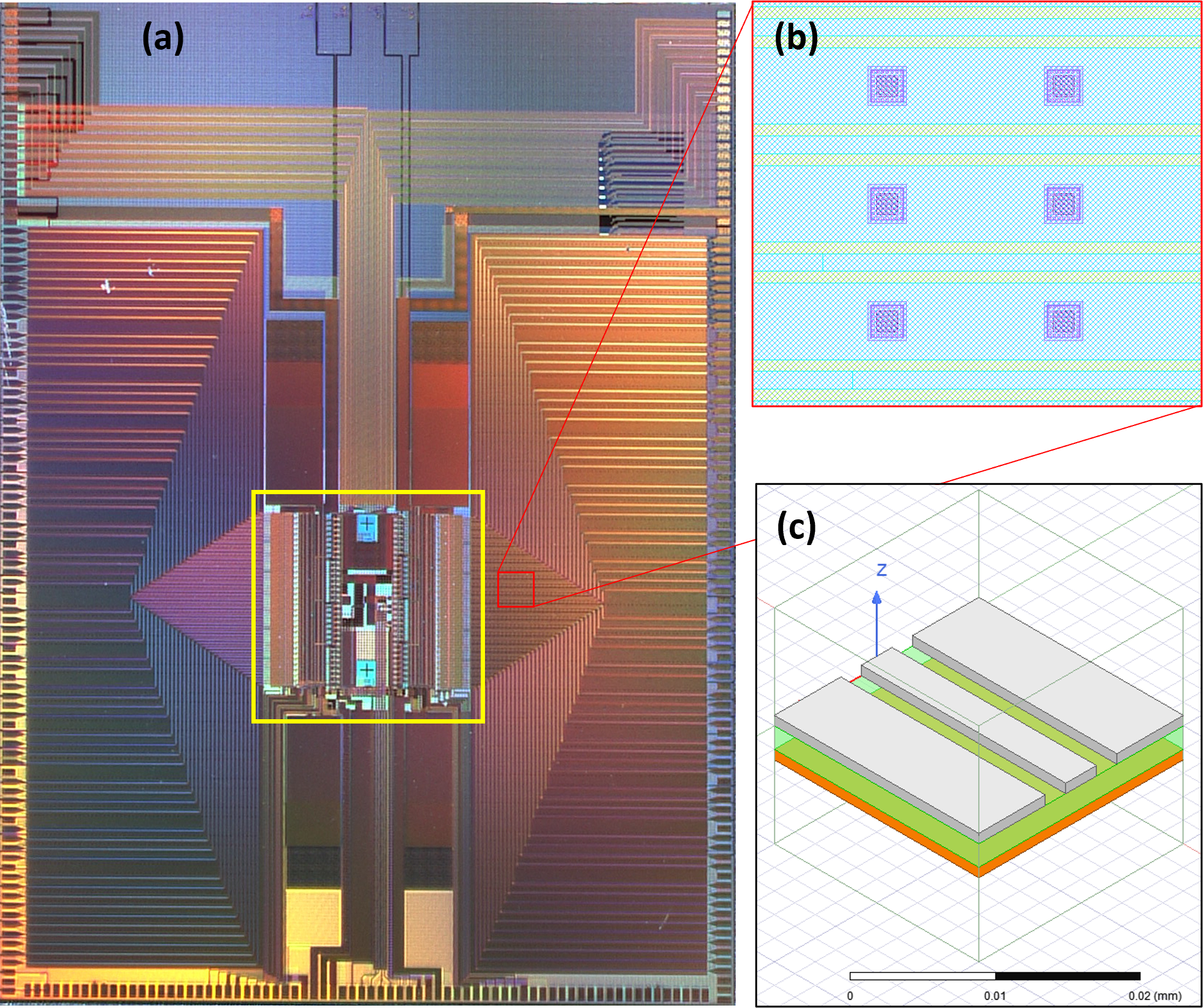}
		\caption{(a) PIC as a 3D/2.5D integration platform. Without TSVs, the PIC was necessary to transition the high-speed data and power/ground signals to wirebond pads. (b) 50$\Omega$ Grounded CPW lines were designed, with ground anchor points located at regular intervals for parasitic mode suppression and better shielding (image from GDSII), (c) The CPW model was designed and simulated in Ansys HFSS.}
		\label{doe_pic}
	\end{figure}

    The architecture contained 32 optical and electrical transceiver channels and 31 data channels with one pair used to optically forward the source clock.  We co-designed the transmitter and receiver circuits in unit cells for WDM system scalability. Optical spectrum planning determined the number of WDM channels and was selected to fit within the free spectral range (FSR) of the optical devices on the same waveguide, ensuring no wavelength reuse. The unit cell array locations were pitch-matched at 40 $\mu$m between the EIC and the PIC. This was selected to be larger than the EIC unit cells' height of 20 $\mu$m shown in Fig. \ref{doe_architecture}, to account for optical routing in the PIC. The EIC was flip-chip 3D integrated on the PIC to achieve the lowest possible interconnect parasitic impedance between them, as wiring parasitics for high-speed signals are critical for optimal performance. The high-speed signal outputs, control signals, and power/ground supplies were routed to wirebond pads at the PIC periphery to be wirebonded to the organic laminate interposer. We used grounded CPW transmission lines for impedance matching the high-speed signal traces, shown in Fig. \ref{doe_pic} (b) and (c). We modeled them in Ansys HFSS to extract the characteristic impedance, and the adjacent lines were separated by a pitch of 6 times the line width to minimize crosstalk below 40dB.
    The EIC was fabricated in GlobalFoundries' 12nm FinFET process, the PIC was fabricated in AIM Photonics' process, and 3D integrated heterogeneously with Direct Bond Interconnect (DBI\textsuperscript{\textregistered}). With a 3D bond interface of approx. $\sim$ 3800 pads, the 3D assembly verification required special techniques to be developed. Even with the dies being co-designed, a final automated verification at the 3D bond interface is crucial. Traditional 2D layout vs. schematic (LVS) verification is unsuitable for verifying 3D structures. The EIC and PIC being co-designed across disparate foundry process development kits (PDKs) posed unique challenges. We needed to verify that the designs matched at the bond interface and avoid any net connectivity mismatch between the EIC-PIC diestack. In partnership with Siemens, we developed a custom 3D LVS workflow to utilize the Xpedition Substrate Integrator (xSI) and the Calibre 3DSTACK tool to exchange EIC and PIC design data, extracting the package-level net locations from GDSII files, aligning with EIC-PIC interface pad geometries, and generating automated mismatch reports after comparing the extracted designs. We reported this process in \cite{siemens2023}. This process allowed us to identify several connectivity issues before finalizing the designs. With the multi-diestack architectures proposed in our platform and the increasing number of bondpads in future implementations, the 3D LVS process is a crucial factor in designing 3D chiplet stack interconnect architectures.
    \begin{figure}
		\centering
		\includegraphics[width=3.4in]{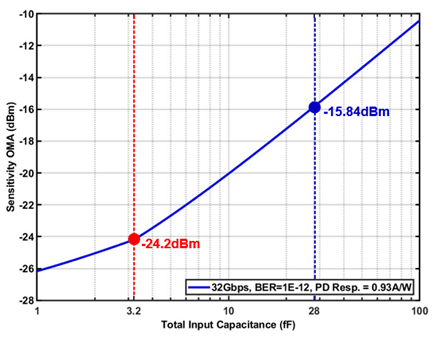}
		\caption{Plot of RX OMA sensitivity vs total TIA RX Input capacitance using parameters from GlobalFoundries GF45SPCLO monolithic electronic-photonic process.}
		\label{rx_sensitivity}
	\end{figure}
    
    To reach $\leq$100\ fJ/bit energy efficiency, we need to further reduce the packaging parasitics between the EIC-PIC pair. In an optical link, the energy efficiency is dominated by the laser power, required to satisfy the receiver sensitivity at the link end, and the receiver deserializer circuit related to the receiver sensitivity. Fig. \ref{rx_sensitivity} presents the simulated receiver optical modulation amplitude (OMA) sensitivity as a function of the total input capacitance of the transimpedance amplifier (TIA), using parameters for the monolithic electronic-photonic GlobalFoundries GF45SPCLO fabless process technology. The photodetector (PD) parameters used in the simulation are sourced from \cite{fu2024}: parasitic capacitance of 0.08fF, dark current of 0.72nA, and responsivity of 0.93 A/W. The simulation incorporates shot noise, a relative-intensity-noise (RIN) of -140dB/Hz, and other receiver noise sources, such as TIA input-referred noise. The targeted operating conditions are a non-return-to-zero (NRZ) data rate of 32 Gb/s at a bit-error-rate (BER) of $10^{-12}$. As mentioned earlier, in \cite{chang2023}, DBI\textsuperscript{\textregistered} provides a superior reduction in input parasitic capacitance compared to C4 and micro-pillar bonding. By monolithically integrating the PD and TIA directly within the GF45SPCLO process, we estimate that the total receiver capacitance can be significantly reduced to 3.2fF. This represents a substantial decrease from the 28fF total input capacitance reported for DBI\textsuperscript{\textregistered} integration in \cite{chang2023}. Consequently, this monolithic integration is projected to improve the OMA sensitivity by 8.36dBm, thereby reducing the required source laser power. Based on a resulting receiver OMA sensitivity of -24.2dBm, we estimate the total source laser power to be 263.95 $\mu$W per channel, accounting for an estimated link loss of 13.98dB, a 2dB link margin, an extinction ratio of 7.7dB, 30$\%$ wall-plug efficiency, and 32 WDM channels. By leveraging projected efficiencies in a contracted CMOS process node transceiver chiplet in our proposed 3D chiplet-stacked electronic-photonic interconnect platform, we provide a pathway to $\leq$100\ fJ/bit energy-efficient, scalable, high-speed data communication, with a large headroom for bandwidth density.

	\section{Future Directions}
	\begin{figure*}
		\centering
		\includegraphics[width=5.0in]{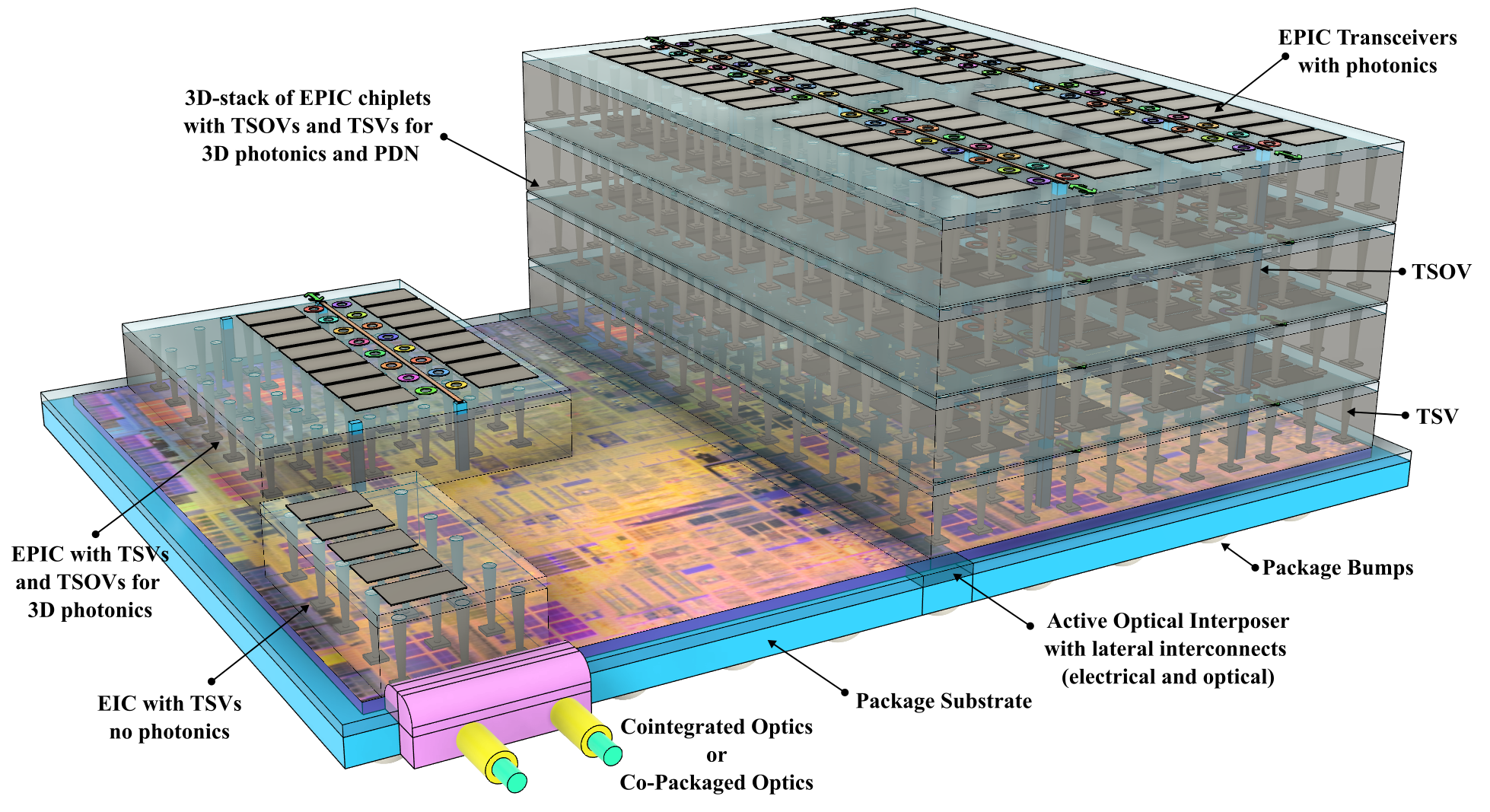}
		\caption{Silicon cityscape of chiplets, with any layer connected to any layer, over global optical interconnect, connected to off-chip via integrated fiber optics.}
		\label{epic_cityscape}
	\end{figure*}
    We are continuing to develop our TSOV fabrication and D2D bonding processes. Future challenges require nanometer-scale alignment in both to achieve low-loss TSOV and airgap-less flipchip bonding. We envision the future of AI HPC architectures driven by a 3D Chiplet stacked electronic-photonic interconnect platform, enabling high-speed, high-bandwidth, energy-efficient, and scalable data communication between any layer in any chiplet stack. Supporting in-package integrated fiber connections for seamless off-package optical communication towards a future where Die-to-Die communication can go global and make true disaggregated computing a reality. With no scalability bottlenecks, this is the key to meeting the demands of next generation of AI.

    \section{Conclusions}
    In this paper, we proposed a 3D chiplet stacked electronic-photonic interconnect platform, which can enable scalable, high-throughput, dense, and energy-efficient interconnect I/O for high-speed communication for optimal applications in future computing systems, in particular, future artificial intelligence systems. We defined the performance metrics of modern interconnects and a figure of merit (FoM) to compare current interconnect standards. We discussed the advantages of optical interconnects over the defined performance metrics, in particular, the energy efficiency and the high-bandwidth density capability of our proposed 3D chiplet stacked electronic-photonic interconnect over the current state-of-the-art 3D electrical interconnect. By benchmarking electrical interconnects against optical interconnects, we presented the decision space for interconnect planning and how 3D photonic interconnects utilizing TSOVs can resolve the current scalability bottlenecks in interconnects. We projected our 3D EPIC platform beating the 3D electrical interconnects to $>$10\ TB/s/$mm^2$ bandwidth density and beyond, with continued development. We reported our development progress on two key physical elements of our proposed interconnect platform, the through-silicon optical via (TSOV), and a flexible die bonding technology in eutectic metallurgy-based 3D integration. We then presented how we leveraged the system-level considerations for 3D electronic-photonic interconnects and implemented a highly scalable, energy-efficient 3D stacked electronic-photonic transceiver system architecture. Lastly, we presented a pathway to extending our demonstrated, industry-ready transceiver design to potentially achieve $\leq$ 100 fJ/bit high-speed communication.
	
	\section*{Acknowledgments}
	This work was supported in part by AFOSR grant No. FA9550-18-1-0186, FA9550-22-1-0532, by ARO grant No. W911NF1910470, by iARPA agreement 2021-21090200004, and by DARPA SRC CHIMES Agreement No. S003605-SRC.

    \begin{IEEEbiographynophoto}{Anirban Samanta}
    received the B.E. degree in Electronics and Communications Engineering from the Birla Institute of Technology, Mesra, India, in 2012 and the M.S. degree in Electrical Engineering from University of California Irvine, Irvine, CA, USA, in 2014. 
    He is currently pursuing the Ph.D. degree in Electrical and Computer Engineering at University of California Davis, Davis, CA. He has previously worked with the ARIANNA Neutrino Detector Group at UC Irvine, developing sensor systems for neutrino detection. His current research interests focuses on 3D electronic-photonic systems co-design, integration, and packaging with applications in high-performance computing.
    \end{IEEEbiographynophoto}
    \begin{IEEEbiographynophoto}{Shun-Hung Lee}
    was born in Taichung, Taiwan in 1995. He received the B.S. degree in electrical engineering from National Yunlin University of Science and Technology, Yunlin, Taiwan, in 2013. He is currently pursuing the M.S. degree in electrical and computer engineering at the University of California, Davis, USA.
    From 2018 to 2019, he was an optical R\&D engineer at Calin Optical Design \& Manufacturing, where he developed optical lenses in collaboration with Waymo and Sony. His current research interest focuses on 3D photonic integration, including Optical TSV and 3D integrated optical devices via ultrafast laser inscription.    
    \end{IEEEbiographynophoto}
    \begin{IEEEbiographynophoto}{Chun-Yi Cheng}
    received the B.S. degree in mechanical engineering from National Central University, Taoyuan, Taiwan, in 2008, and the M.S. degree in the Institute of Electrical and Control Engineering from National Yang Ming Chiao Tung University, Hsinchu, Taiwan, in 2010. He is currently working toward the Ph.D. degree in electrical engineering with Texas A\&M University, College Station, TX, USA. From 2011 to 2014, he was a senior engineer with Faraday Technology Corporation, Hsinchu, Taiwan, From 2016 to 2017, he was an analog deputy project manager with Fresco Logic, Taipei, Taiwan. From 2017 to 2021, he was a senior analog design engineer with Silicon Motion Technology. In 2024 and 2025, he was a research intern with IBM, Yorktown Heights, NY, USA. His research interests include design of high-speed circuits for electrical and optical wireline communication.
    \end{IEEEbiographynophoto}
    \begin{IEEEbiographynophoto}{Samuel Palermo}
    received the B.S. and M.S. degrees in electrical engineering from Texas A\&M University, College Station, TX in 1997 and 1999, respectively, and the Ph.D. degree in electrical engineering from Stanford University, Stanford, CA in 2007. 

    From 1999 to 2000, he was with Texas Instruments, Dallas, TX, where he worked on the design of mixed-signal integrated circuits for high-speed serial data communication. From 2006 to 2008, he was with Intel Corporation, Hillsboro, OR, where he worked on high-speed optical and electrical I/O architectures. In 2009, he joined the Electrical and Computer Engineering Department of Texas A\&M University where he is currently the J. W. Runyon Jr. Professor. His research interests include high-speed electrical and optical interconnect architectures, RF photonics, radiation-hardened electronics, and AI computing hardware. 

    Dr. Palermo is a recipient of a 2013 NSF-CAREER award. He is a member of Eta Kappa Nu and IEEE. He is currently an associate editor for IEEE Journal of Solid-State Circuits and has previously served in this role for IEEE Solid-State Circuits Letters and IEEE Transactions on Circuits and System – II. He has also previously served as a distinguished lecturer for the IEEE Solid-State Circuits Society and on the IEEE CASS Board of Governors. He was a coauthor of the Jack Raper Award for Outstanding Technology-Directions Paper at the 2009 International Solid-State Circuits Conference, the Best Student Paper at the 2014 Midwest Symposium on Circuits and Systems, an Outstanding Student Paper Award at the 2018 Custom Integrated Circuits Conference, and the Best Student Paper Award at the 2024 Opto-Electronics and Communications Conference. He received the Texas A\&M University Department of Electrical and Computer Engineering Outstanding Professor Award in 2014 and the Engineering Faculty Fellow Award in 2015.
    \end{IEEEbiographynophoto}
    \begin{IEEEbiographynophoto}{S. J. Ben Yoo}
     [S’82, M’84, SM’97, F’07] (Fellow of IEEE and Fellow of Optica) is currently a Distinguished Professor with UC Davis.  He is currently leading the 3D EPIC AI Project and the UC Davis part of Northwest-AI Microelectronics Commons Hub under the US CHIPS and Science Act, the ExPlor Project under the US Air Force Office of Scientific Research, the NanoHybrid project Photonic Processing Activity under US Department of Energy’s Micro Microelectronics Science Research Center, and the NaPSAC project under DARPA support.  His research at UC Davis includes 2D/3D photonic integration for future computing, cognitive networks, communication, imaging, and navigation systems, micro/nano systems integration, and the future Internet. Prior to joining UC Davis in 1999, he was a Senior Research Scientist at Bellcore, leading technical efforts in integrated photonics, optical networking, and systems integration. His research activities at Bellcore included the next-generation Internet, reconfigurable multiwavelength optical networks (MONET), wavelength interchanging cross connects, wavelength converters, vertical-cavity lasers, and high-speed modulators. He led the MONET testbed experimentation efforts, and participated in ATD/MONET systems integration and a number of standardization activities.  Prior to joining Bellcore in 1991, he conducted research on nonlinear optical processes in quantum wells, a four-wave-mixing study of relaxation mechanisms in dye molecules, and ultrafast diffusion-driven photodetectors at Stanford University (BS’84, MS’86, PhD’91, Stanford University). Prof. Yoo is Fellow of IEEE, OPTICA, NIAC and a recipient of the DARPA Award for Sustained Excellence, the Bellcore CEO Award, the Mid-Career Research Faculty Award (UC Davis), the Senior Research Faculty Award (UC Davis), and numerous best paper awards from IEEE, ACM, and Optica conferences.
    \end{IEEEbiographynophoto}
	

\begin{thebibliography}{1}
		\bibliographystyle{IEEEtran}
		
		\bibitem{epoch_models} Robi Rahman, David Owen and Josh You (2024), "Tracking Large-Scale AI Models". Published online at epoch.ai. Retrieved from: 'https://epoch.ai/blog/tracking-large-scale-ai-models' [online resource]
		
		\bibitem{epoch_scaling_2030} Jaime Sevilla et al. (2024), "Can AI Scaling Continue Through 2030?". Published online at epoch.ai. Retrieved from: 'https://epoch.ai/blog/can-ai-scaling-continue-through-2030' [online resource]

        \bibitem{duang2021} G. Duan, Y. Kanaoka, R. McRee, B. Nie and R. Manepalli, "Die Embedding Challenges for EMIB Advanced Packaging Technology," 2021 IEEE 71st Electronic Components and Technology Conference (ECTC), San Diego, CA, USA, 2021, pp. 1-7.
		
		\bibitem{Naeemi2004} Naeemi, Azad, Jianping Xu, T. K. Gaylord, and J. D. Meindl. "Optical and electrical interconnect partition length based on chip-to-chip bandwidth maximization." IEEE Photonics Technology Letters 16, no. 4 (2004): 1221-1223.
		
		\bibitem{beausoleil2008} Beausoleil, Raymond G., Philip J. Kuekes, Gregory S. Snider, Shih-Yuan Wang, and R. Stanley Williams. "Nanoelectronic and nanophotonic interconnect." Proceedings of the IEEE 96, no. 2 (2008): 230-247.
		
		\bibitem{stucchi2013} Stucchi, Michele, Stefan Cosemans, Joris Van Campenhout, Zsolt Tőkei, and Gerald Beyer. "On-chip optical interconnects versus electrical interconnects for high-performance applications." Microelectronic engineering 112 (2013): 84-91.
		
		\bibitem{gholami2024} Gholami, Amir, Zhewei Yao, Sehoon Kim, Coleman Hooper, Michael W. Mahoney, and Kurt Keutzer. "AI and memory wall." IEEE Micro (2024).
		
		\bibitem{kim2021} Kim, Seongguk, Subin Kim, Kyungjun Cho, Taein Shin, Hyunwook Park, Daehwan Lho, Shinyoung Park et al. "Signal integrity and computing performance analysis of a processing-in-memory of high bandwidth memory (PIM-HBM) scheme." IEEE Transactions on Components, Packaging and Manufacturing Technology 11, no. 11 (2021): 1955-1970.
		
		\bibitem{kim2024} Kim, Kwiwook, and Myeong-jae Park. "Present and future, challenges of high bandwith memory (HBM)." In 2024 IEEE International Memory Workshop (IMW), pp. 1-4. IEEE, 2024.
		
		\bibitem{zhou-micron-2023} Zhou, Wei, Michael Kwon, Yingta Chiu, Huimin Guo, Bharat Bhushan, Bret Street, Kunal Parekh, and Akshay Singh. "Critical challenges with copper hybrid bonding for chip-to-wafer memory stacking." In 2023 IEEE 73rd Electronic Components and Technology Conference (ECTC), pp. 336-341. IEEE, 2023.		
		
		\bibitem{naffziger-AMD-2021} Naffziger, Samuel, Noah Beck, Thomas Burd, Kevin Lepak, Gabriel H. Loh, Mahesh Subramony, and Sean White. "Pioneering chiplet technology and design for the amd epyc™ and ryzen™ processor families: Industrial product." In 2021 ACM/IEEE 48th Annual International Symposium on Computer Architecture (ISCA), pp. 57-70. IEEE, 2021.
		
		\bibitem{loh-AMD-2023} Loh, Gabriel H., and Raja Swaminathan. "The next era for chiplet innovation." In 2023 Design, Automation \& Test in Europe Conference \& Exhibition (DATE), pp. 1-6. IEEE, 2023.
		
		\bibitem{ucie2024} Das Sharma, D., Pasdast, G., Tiagaraj, S. et al. High-performance, power-efficient three-dimensional system-in-package designs with universal chiplet interconnect express. Nat Electron 7, 244–254 (2024). https://doi.org/10.1038/s41928-024-01126-y
		
		\bibitem{elsherbini-intel-2021} Elsherbini, Adel, Shawna Liff, Johanna Swan, Kimin Jun, Sathya Tiagaraj, and Gerald Pasdast. "Hybrid bonding interconnect for advanced heterogeneously integrated processors." In 2021 IEEE 71st Electronic Components and Technology Conference (ECTC), pp. 1014-1019. IEEE, 2021.
		
		\bibitem{ghose2018} Boroumand, Amirali, Saugata Ghose, Youngsok Kim, Rachata Ausavarungnirun, Eric Shiu, Rahul Thakur, Daehyun Kim et al. "Google workloads for consumer devices: Mitigating data movement bottlenecks." In Proceedings of the twenty-third international conference on architectural support for programming languages and operating systems, pp. 316-331. 2018.
		
		\bibitem{dally2011} S. W. Keckler, W. J. Dally, B. Khailany, M. Garland and D. Glasco, "GPUs and the Future of Parallel Computing," in IEEE Micro, vol. 31, no. 5, pp. 7-17, Sept.-Oct. 2011, doi: 10.1109/MM.2011.89.
		
		\bibitem{keeler2018} DARPA Photonics in the Package for Extreme Scalability (PIPES), 2018
		
		\bibitem{gf45spclo} https://gf.com/technology-platforms/silicon-photonics/
		
		\bibitem{amdgenoa} https://www.tomshardware.com/reviews/amd-4th-gen-epyc-genoa-9654-9554-and-9374f-review-96-cores-zen-4-and-5nm-disrupt-the-data-center
		
		\bibitem{cowos2023} Y. -C. Hu et al., "CoWoS Architecture Evolution for Next Generation HPC on 2.5D System in Package," 2023 IEEE 73rd Electronic Components and Technology Conference (ECTC), Orlando, FL, USA, 2023, pp. 1022-1026, doi: 10.1109/ECTC51909.2023.00174.
		
		\bibitem{foveros2022} W. Gomes et al., "Ponte Vecchio: A Multi-Tile 3D Stacked Processor for Exascale Computing," 2022 IEEE International Solid-State Circuits Conference (ISSCC), San Francisco, CA, USA, 2022, pp. 42-44, doi: 10.1109/ISSCC42614.2022.9731673.

        \bibitem{netherton2024} Netherton, A., Dumont, M., Nelson, Z., Koo, J., Jhonsa, J., Mo, A., Bowers, J. E. (2024). High capacity, low power, short reach integrated silicon photonic interconnects. Photonics Research, 12(11), A69-A86.

        \bibitem{chen2014} Chen, Y., Mak, P. I., Zhang, L., Wang, Y. (2014). A 0.002-mm$^{2} $6.4-mW 10-Gb/s Full-Rate Direct DFE Receiver With 59.6$\%$ Horizontal Eye Opening Under 23.3-dB Channel Loss at Nyquist Frequency. IEEE Transactions on Microwave Theory and Techniques, 62(12), 3107-3117.

        \bibitem{manian2017} Manian, A., Razavi, B. (2017). A 40-Gb/s 14-mW CMOS wireline receiver. IEEE Journal of Solid-State Circuits, 52(9), 2407-2421.

        \bibitem{kim2018} D. Kim, W. -S. Choi, A. Elkholy, J. Kenney and P. K. Hanumolu, "A 15Gb/s 1.9pJ/bit sub-baud-rate digital CDR," 2018 IEEE Custom Integrated Circuits Conference (CICC), San Diego, CA, USA, 2018, pp. 1-4.

        \bibitem{liu2021} J. Liu, G. Huang, R.N. Wang, J. He, A.S. Raja, T. Liu, N.J. Engelsen, and T.J. Kippenberg, "High-yield, wafer-scale fabrication of ultralow-loss, dispersion-engineered silicon nitride photonic circuits," Nat. Commun. 12, 2236 (2021).

        \bibitem{ucie2022} D. Das Sharma, G. Pasdast, Z. Qian and K. Aygun, "Universal Chiplet Interconnect Express (UCIe): An Open Industry Standard for Innovations With Chiplets at Package Level," in IEEE Transactions on Components, Packaging and Manufacturing Technology, vol. 12, no. 9, pp. 1423-1431, Sept. 2022.

        \bibitem{fan2024} Y. Fan, H. Gan, Y. Zhou, B. Lei, G. Song and Q. Wang, "Signal Integrity Simulation and Analysis for 2.5D Advanced Package Interconnect Based on Universal Chiplet Interconnect Express (UCIe)," 2024 IEEE 26th Electronics Packaging Technology Conference (EPTC), Singapore, 2024, pp. 401-406.

        \bibitem{zhang2018} Y. Zhang, Y. -C. Ling, Y. Zhang, K. Shang and S. J. B. Yoo, "High-Density Wafer-Scale 3-D Silicon-Photonic Integrated Circuits," in IEEE Journal of Selected Topics in Quantum Electronics, vol. 24, no. 6, pp. 1-10, Nov.-Dec. 2018.

        \bibitem{anirban2023} A. Samanta et al., "A Direct Bond Interconnect 3D Co-Integrated Silicon-Photonic Transceiver in 12nm FinFET with -20.3dBm OMA Sensitivity and 691fJ/bit," 2023 Optical Fiber Communications Conference and Exhibition (OFC), San Diego, CA, USA, 2023, pp. 1-3, doi: 10.1364/OFC.2023.M3I.4.

        \bibitem{chang2023} P. -H. Chang et al., "A 3D Integrated Energy-Efficient Transceiver Realized by Direct Bond Interconnect of Co-Designed 12 nm FinFET and Silicon Photonic Integrated Circuits," in Journal of Lightwave Technology, vol. 41, no. 21, pp. 6741-6755, 1 Nov.1, 2023.

        \bibitem{fu2024} M. Fu and S. J. Ben Yoo, "0.08 fF, 0.72 nA dark current, 91$\%$ Quantum Efficiency, 38 Gb/s Nano-photodetector on a 45 nm CMOS Silicon-Photonic Platform," 2024 Optical Fiber Communications Conference and Exhibition (OFC), San Diego, CA, USA, 2024, pp. 1-3.

        \bibitem{siemens2023} https://resources.sw.siemens.com/en-US/white-paper-taking-chips-to-another-dimension-3d-integrated-silicon-photonic-transceiver/

        \bibitem{dbi2020} G. Gao et al., "Die to Wafer Hybrid Bonding: Multi-Die Stacking with Tsv Integration," 2020 International Wafer Level Packaging Conference (IWLPC), San Jose, CA, USA, 2020, pp. 1-8.
		
	\end{thebibliography}
\end{document}